\documentclass[reprint,aps,amsmath,amssymb,twoside,prd,showkeys,superscriptaddress,nofootinbib]{revtex4-1}
\usepackage{iftex}
\ifPDFTeX
    \pdfoutput=1
\fi

\usepackage{mathrsfs}
\usepackage[bottom]{footmisc}
\usepackage{enumitem}
\usepackage{booktabs}

\usepackage{tabularx}
\usepackage{cancel}
\usepackage{pifont}
\usepackage{soul}
\usepackage[colorlinks=true,linktocpage=true,linkcolor=blue,citecolor=blue,urlcolor=blue]{hyperref}

\usepackage{mathtools,amsmath,amssymb,mathrsfs,color,esint}
\usepackage{array}
\usepackage[dvipsnames]{xcolor}
\newcolumntype{C}[1]{>{\centering\arraybackslash}p{#1}}

\allowdisplaybreaks
 
\begin{document}

\title{{\bf The Gregory-Laflamme Instability of five-dimensional Black Strings in Einstein-Gauss-Bonnet Gravity}}

\author{Pau Figueras}
\author{\'Aron D. Kov\'acs}
\author{Shunhui Yao}

\affiliation{Centre for Geometry, Analysis and Gravitation, School of Mathematical Sciences, Queen Mary University of London, Mile End Road, London E1 4NS, United Kingdom}

\date{\today}

\begin{abstract}
We complement and extend our recent study of the non-linear evolution of the Gregory-Laflamme instability of black strings in Einstein-Gauss-Bonnet gravity in five spacetime dimensions. First, we study time-dependent perturbations of uniform black strings in Einstein-Gauss-Bonnet theory and determine the growth rates as a function of the coupling. We find that for positive Gauss-Bonnet couplings, the black strings are more unstable than in general relativity in the sense that the growth rates are larger and shorter wavelengths are unstable. In the second part of the paper we show that the hyperbolicity of the equation of motion of Einstein-Gauss-Bonnet theory is determined by an effective metric when the dynamics is restricted to spherical symmetry in the asymptotically flat directions. When the effective field theory becomes strongly coupled, as it happens near the endpoint of the Gregory-Laflamme instability of the black string, this effective metric becomes degenerate and the Cauchy evolution breaks down. Applying this criterion to uniform black strings, we find that for any positive Gauss-Bonnet coupling there is an open set of regular solutions (including the one of minimal thickness) on which the effective metric degenerates in a neighbourhood of the horizon. Therefore, we argue that these solutions cannot be attained from weakly coupled initial data by Cauchy evolution and the solutions of the linearized equations around them cannot be interpreted as approximations to nearby non-linear evolutions. We provide further evidence for this picture with non-linear simulations of the Gregory-Laflamme instability. The non-linear simulations terminate through a loss of hyperbolicity of the Tricomi or Keldysh type, depending on the sign of the Gauss-Bonnet coupling.

\end{abstract}

\maketitle
\section{Introduction}
\label{sec:Intro}

The Gregory-Laflamme (GL) instability of black strings (and $p$-branes in general) \cite{Gregory:1993vy,Gregory:1994bj} has played a pivotal role in the modern understanding of general relativity (GR) and, in particular, of its strong field dynamics in dimensions higher than four. The reason is that it showed that (non-extremal) vacuum black holes in higher dimensions can be unstable, which  is in stark contrast with the four dimensional case, where only recently the (sub-extremal) Kerr black hole family has been shown to be non-linearly stable \cite{Hintz:2026ami,Szeftel:2026qzr}.

The GL instability, first discovered for black $p$-branes in Kaluza-Klein (KK) spacetimes, is generic in higher dimensional spacetimes in the sense that it can also affect  black holes in asymptotically flat (AF) or anti-de Sitter (AAdS) spacetimes. The reason is that, unlike in four dimensions where the spin of a Kerr black hole is bounded, in dimensions greater than five black holes can rotate arbitrarily fast without ever reaching an extremal limit. In this rapidly rotating regime,   \cite{Emparan:2003sy} recognised that, at least locally, such black holes resembled a certain $p$-brane and hence they ought to be unstable under a GL-like instability. This expectation was later confirmed in a variety of situations \cite{Dias:2009iu,Dias:2010eu,Dias:2010gk,Santos:2015iua}. Furthermore, even though some black holes may not initially be unstable under the GL instability, the non-linear dynamics of GR may drive them into a regime where such instabilities locally develop \cite{Bantilan:2019bvf,Andrade:2020dgc}, thus showing that this instability is indeed a generic feature of GR in higher dimensions. 

In this rapidly spinning regime of higher dimensional black holes, the horizon is typically described by widely separated length scales. \cite{Emparan:2009cs} exploited this property to devise a perturbative scheme to construct new families of higher dimensional stationary black holes, with new types of horizon topologies that do not have analogues in four dimensions. Therefore, the GL instability of black strings together with the discovery of the black ring \cite{Emparan:2001wn} have driven research in GR beyond the traditional astrophysical settings and brought important new insights into Einstein's theory of gravity.

Given the ubiquity of the GL instability in higher dimensional GR, it is natural to enquire about its endpoint. \cite{Gregory:1993vy} conjectured, based on entropic arguments, that the endpoint of GL instability of black strings would be a collection of spherical black holes aligned along the KK direction. However, the issue with this picture is that horizons cannot smoothly bifurcate (see e.g., \cite{Hawking:1973uf}) and hence, a naked singularity would have to appear where the string pinches off, thus potentially violating the weak cosmic censorship (WCC) conjecture in higher dimensions.  Evidence for such violations was first found in black strings \cite{Lehner:2010pn} (see also the more recent work \cite{Figueras:2022zkg}) and later extended to AF rotating black holes and black rings \cite{Figueras:2015hkb,Figueras:2017zwa,Bantilan:2019bvf}.

Near the endpoint of the GL instability, regions of large curvature develop on and outside the horizon. Taking the point of view of GR as a low energy effective field theory (EFT) of gravity, one would expect that the higher derivative corrections to the Einstein-Hilbert term in the action become important in such regions and can alter the local dynamics. In our companion paper \cite{Figueras:2026epx} we studied this question focusing on  the effects on the dynamics of the GL instability of black strings in five spacetime dimensions of the Gauss-Bonnet term, which is the leading order higher derivative correction to vacuum GR: 
\begin{equation}
    S=\frac{1}{16\pi G}\int d^5x\sqrt{-g}\left(R+\lambda^\text{GB}\,\mathscr{L}_\text{GB}\right)\,, \label{eq:action5d}
\end{equation}
where $\lambda^\text{GB}$ is the (dimensionful) coupling constant and $\mathscr{L}_{GB}$ is the Gauss-Bonnet invariant:
\begin{equation}
    \mathscr{L}_\text{GB}\equiv R^2 -4\,R_{\mu\nu}\,R^{\mu\nu}+R_{\mu\nu\rho\sigma}\,R^{\mu\nu\rho\sigma}\,.
\end{equation}
We refer to the theory \eqref{eq:action5d} as Einstein-Gauss-Bonnet (EGB) gravity. 

The equations of motion that follow from varying \eqref{eq:action5d} with respect to the spacetime metric $g_{\mu\nu}$  are given by
\begin{equation}
    R_{\mu\nu} + \lambda^\text{GB}\,H_{\mu\nu}=0, \label{eq:eomsEGB}
\end{equation}
where
\begin{equation}
\begin{gathered}
     H_{\mu\nu} = 2\,R\,R_{\mu\nu}-4\,R_\mu^{\phantom{\mu}\rho}\,R_{\nu\rho}-4\,R^{\rho\sigma}\,R_{\mu\rho\nu\sigma}
     \\+2\,R_\mu^{\phantom{\mu}\rho\sigma\tau}\,R_{\nu\rho\sigma\tau}-\tfrac{1}{3}\,g_{\mu\nu}\,\mathscr{L}_{GB}\,. \label{eq:def_H}
\end{gathered}
\end{equation}  

The truncated theory \eqref{eq:action5d} has a well-posed initial value problem \cite{Kovacs:2020ywu,AresteSalo:2023hcp,Figueras:2024bba,SY_paper} for sufficiently small values of the coupling $|\lambda^\text{GB}|$ and hence it is a consistent classical EFT. However, consistency with an underlying UV complete theory of gravity selects the positive sign on $\lambda^\text{GB}$ \cite{Bellazzini:2015cra,Cheung:2016wjt}.

For $\lambda^\text{GB}>0$, \cite{Figueras:2026epx} found that the GL cascade develops faster than in GR and that the classical evolution terminates when the thickness of the thinnest string-like segments becomes $O(\sqrt{\lambda^\text{GB}})$, i.e., of the order of the UV cutoff length scale. It is tempting to relate this to the fact that for $\lambda^\text{GB}>0$ UBSs in EGB gravity have a minimum thickness $r_H\geq\sqrt{8\lambda^\text{GB}}$, and that the limiting UBS is regular \cite{Kobayashi:2004hq,Suranyi:2008wc,Brihaye:2010me}. This would suggest that the thinnest segments of the string might locally settle down to this limiting solution, so that curvatures remain bounded. However, as we show in this paper, the evolution does not approach this solution. Instead, it terminates through a loss of hyperbolicity of the equations of motion, while the thinnest segments are still thicker than $\sqrt{8\lambda^\text{GB}}$ and still contracting. In fact, UBSs that are close enough to the minimum thickness lie outside the regime in which the initial value problem of EGB gravity is well-posed, and hence they cannot be reached in a Cauchy evolution. Once the thinnest portions of the string become $O(\sqrt{\lambda^\text{GB}})$, stringy effects presumably take over the local dynamics and implement the topology change as suggested in \cite{Emparan:2024mbp}.

On the other hand, for  $\lambda^\text{GB}<0$ the evolution of the GL instability is qualitatively similar to GR, and there does not seem to be any mechanism preventing the black string from developing arbitrarily large curvatures on and outside the horizon in finite asymptotic time. In this case there is a lower bound on the thickness of a UBS solution (for fixed Gauss-Bonnet coupling).  However, as this lower bound approaches, curvature invariants on the horizon of a UBS can take arbitrarily large values and the entropy tends to zero. In the non-linear evolution of the black string in \cite{Figueras:2026epx} we can also observe a more significant growth of curvature invariants on and outside the apparent horizon than in the $\lambda^\text{GB}>0$ case. 

For either sign of $\lambda^\text{GB}$, the Cauchy evolution breaks down before reaching the endpoint of the instability. This happens when the diagnostics of Section~\ref{subsec:hypberb} signal a loss of hyperbolicity in a region inside the apparent horizon (AH) near the thinnest parts of the string. Since we use horizon-penetrating coordinates, our evolutions cannot be continued beyond this point. The exterior evolution could in principle be continued further, e.g., by excising the region where hyperbolicity is lost. However, for positive couplings a continuation cannot lead to the near-threshold UBSs discussed above, since a degeneracy of the effective metric would first have to form outside the black hole. Note that for $\lambda^\text{GB}<0$ the EFT description also breaks down due to the growth of curvature, whereas in the $\lambda^\text{GB}>0$ case the 6-derivative EFT corrections are still negligible in the black string exterior. 

As summarised above, the focus of   \cite{Figueras:2026epx} is on the non-linear evolution of the GL instability of black strings in EGB gravity and how a more fundamental theory might prevent the formation of a naked singularity. This paper complements and extends the results in \cite{Figueras:2026epx} in three ways. In the first part, we provide a detailed description of the relevant properties of uniform black strings in EGB gravity and study their linear (in)stability as a function of the coupling. In the second part, we study the hyperbolicity of the equations of motion with an $SO(3)$ symmetry along the asymptotically flat directions. This is the relevant isometry group for the spacetime metric in the non-linear evolution of the GL instability of black strings in \cite{Figueras:2026epx}. It turns out that in this symmetry-reduced setting only a single the physical degree of freedom is dynamical and its propagation is described by an effective metric. When the EFT becomes strongly coupled, this effective metric can become degenerate, leading to a loss of hyperbolicity of the equations regardless of gauge choice. Therefore, this effective metric provides a useful diagnostic. In the third part, we present some details of the non-linear numerical simulations of the GL instability, complementing the results of \cite{Figueras:2026epx}.

\subsection{Summary of the main results}

Uniform black strings (UBSs) in EGB theory were first constructed \cite{Kobayashi:2004hq}, see also \cite{Suranyi:2008wc} and \cite{Brihaye:2010me}. In this paper we reproduce their results using modern numerical methods and compute the thermodynamic quantities of the UBS; in particular, the Wald entropy (top panel in Fig. \ref{fig:physQ_UBS}).\footnote{Ref. \cite{Suranyi:2008wc} computes the entropy of the UBS in EGB gravity by integrating the First Law.} 

In five spacetime dimensions, UBSs in EGB theory only exist within a range of couplings:\footnote{Equivalently, this bound implies that UBSs in EGB have a minimum thickness for a given (positive) coupling $\lambda^\text{GB}$, namely $r_H\geq\sqrt{8\lambda^\text{GB}}$. As we shall argue in Section~\ref{subsec:hypberb}, however, UBSs that are close enough to this minimum thickness cannot be attained dynamically.}
\begin{equation}
    -\frac{1}{4}< \frac{\lambda^\text{GB}}{r_H^2}\leq \frac{1}{8}\,,
    \label{eq:range_UBS}
\end{equation}
where $r_H$ is the radius of the horizon $S^2$, and it is a gauge invariant quantity in the symmetry class of spacetimes that we consider. It is interesting to note that the tension of the string depends on the sign of $\lambda^\text{GB}$, with the tension monotonically decreasing as  $\lambda^\text{GB}$ becomes more positive, and increasing as $\lambda^\text{GB}$ becomes more negative, see the top panel in Fig. \ref{fig:physQ_UBS}. As we shall see shortly, the behavior of the tension as a function of the coupling is correlated with the stability properties of UBSs.

We charaterize the geometry of the horizon of the UBS by evaluating suitably normalized quadratic and cubic curvature invariants on the horizon, see eqs. \eqref{eq:norm_K} and \eqref{eq:norm_S}. The normalization has been chosen so that both evaluate to $1$ for a UBS in GR; in EGB theory, however, these curvature invariants are functions of $\lambda^\text{GB}/r_H^2$. The bottom panel in Fig. \ref{fig:physQ_UBS} shows that the curvature invariants diverge and the entropy goes to zero as $\lambda^\text{GB}/r_H^2\rightarrow -\frac{1}{4}$, while these quantities remain finite as $\lambda^\text{GB}/r_H^2\rightarrow \frac{1}{8}$. Therefore, the solution reached in the limit $\lambda^\text{GB}/r_H^2\rightarrow -\frac{1}{4}$ is a naked singularity, with divergent curvature polynomials, while the solution with $\lambda^\text{GB}/r_H^2= \frac{1}{8}$ has finite entropy and thickness $r_H= \sqrt{8\lambda^\text{GB}}$, and finite curvature invariants on the horizon. However, as we shall see in Section~\ref{subsec:hypberb}, UBSs with $\tfrac14(\sqrt2-1)<\lambda^\text{GB}/r_H^2\leq\tfrac18$ lie outside the regime in which EGB gravity has a well-posed initial value problem. Therefore, even though they are regular solutions (on and outside the horizon) of the equations of motion, we argue that they cannot be attained dynamically starting from weakly coupled initial data.

The values of the curvature invariants on the horizon shown in the bottom panel of Fig. \ref{fig:physQ_UBS} as functions of $\lambda^\text{GB}/r_H^2$ are useful to characterize the local string segments in the evolution of the GL instability in \cite{Figueras:2026epx}. Recall that in GR, the evolution of GL instability of black strings results in a dynamical horizon that can be well-approximated by a sequence of quasi-stationary spherical black holes joined by uniform black strings on different scales, and the rapid dynamics is restricted to the thinnest regions where the strings join the black holes. Evidence for this picture comes from the fact that curvature invariants \eqref{eq:norm_K}--\eqref{eq:norm_S} evaluated on the apparent horizon (AH) of dynamical string  match, very accurately, those of static spherical black holes and black strings. On the other hand, in EGB theory, \cite{Figueras:2026epx} showed that on the uniform string-like segments of the AH of the unstable black string, the curvature invariants vary on a scale comparable to the length of the segments. Therefore, one cannot accurately describe these segments as quasi-stationary UBSs with local coupling $\lambda^\text{GB}/r_H^2$. 

In Section \ref{sec:Linear_analysis} we study the linear stability of UBS in EGB theory under perturbations that preserve the spherical symmetry (along the asymptotically flat directions) of the background. We first study the zero modes, namely time-independent modes, that signal the onset of the GL instability and our results agree with those in \cite{Suranyi:2008wc}, see Fig. \ref{fig:zero_modes}. As this figure shows, the critical wave number $k_c\,r_H$ increases as $\lambda^\text{GB}/r_H^2$ increases, and it sharply increases in the limit $\lambda^\text{GB}/r_H^2\to \frac{1}{8}$. On the other hand, $k_c\,r_H$  decreases as $\lambda^\text{GB}/r_H^2$ becomes more negative, indicating that, for $\lambda^\text{GB}/r_H^2<0$, thinner strings are more stable in EGB  than in GR. However, for $\lambda^\text{GB}/r_H^2\lesssim -0.23$, the critical wave number $k_c\,r_H$ starts increasing again as $\lambda^\text{GB}/r_H^2\to -\frac{1}{4}$. This should not be surprising since the background becomes singular in this limit. However, we emphasise again that for $\lambda^\text{GB}/r_H^2>\tfrac14(\sqrt2-1)$ these zero modes are static solutions of linearized equations that fail to be hyperbolic near the horizon. Therefore, in this range they should not be interpreted as signalling the onset of a genuine dynamical instability.

The study of the time-dependent perturbations of black strings in EGB theory in five spacetime dimensions presented here extends previous results of \cite{Henriquez-Baez:2022bfi} using perturbation theory in the Gauss-Bonnet coupling (see also results of \cite{Giacomini:2015dwa} on seven dimensional black strings). The growth rates of the GL instabilities for some representative values of $\lambda^\text{GB}/r_H^2$ are shown in Fig. \ref{fig:growth_rates}.\footnote{We only consider UBS backgrounds below the hyperbolicity threshold $\lambda^\text{GB}/r_H^2<\tfrac14(\sqrt2-1)$. Beyond this regime the linearized initial value problem on the UBS background is ill-posed and growth rates (though they may be computed) are not meaningful.} We find that the growth rates follow the same trend as the zero modes: for increasingly positive values of $\lambda^\text{GB}/r_H^2$, the growth rate $\Omega\,r_H$ also grows, confirming that these strings are more unstable. On the other hand, $\Omega\,r_H$ decreases for negative values of the coupling until it starts growing again $\lambda^\text{GB}/r_H\lesssim -0.23$, as the background becomes more singular. This behavior of the unstable modes is intuitive from the tension of the strings as a function of the coupling.

In Section \ref{sec:mccz4_cartoon} we study the well-posedness of the evolution  equations EGB gravity in the mCCZ4 formulation \cite{AresteSalo:2022hua,AresteSalo:2023mmd} (see also \cite{Kovacs:2020pns,Kovacs:2020pns}), specializing to spacetimes within the symmetry class that is relevant for the evolution of the GL instability of black strings, namely $SO(3)$ symmetry along the asymptotically flat directions (see Appendix \ref{app:evolution_equations}). This symmetry is implemented at the level of the equations of motion using the modified cartoon method \cite{Pretorius:2004jg,Cook:2016soy}. By analysing the principal symbol of the symmetry-reduced equations of motion, we identify the effective metric that governs the high frequency propagation of the physical degree of freedom, eq. \eqref{eq:eff_metric}. This metric differs from the spacetime metric by spacetime curvature terms that are proportional to $\lambda^\text{GB}$. Therefore, when the EFT becomes strongly coupled, this effective metric can have Euclidean signature, implying that the equation of motion is elliptic and the corresponding initial value problem is ill-posed. 

In Section \ref{subsec:hypberb} we develop diagnostic of the hyperbolicity loss in terms of the eigenvalues of the spatially reduced effective metric \eqref{eq:spatial_eff_metric}, which follows from \eqref{eq:eff_metric}. Evaluating this diagnostic on the UBSs of Section~\ref{sec:UBS_EGB}, we find that the effective metric on the horizon ceases to be Lorentzian for
\begin{equation*}
\tfrac14(\sqrt2-1)<\tfrac{\lambda^\text{GB}}{r_H^2}\leq\tfrac18\,,
\end{equation*}
and, by continuity, the same holds in a neighbourhood of the horizon, including a region in the black string exterior. This suggests that such UBSs, as well as the solutions of the linearized equations around them, have no dynamical relevance since they cannot be reached by Cauchy evolution. On the other hand, for $\lambda^\text{GB}<0$ the symmetry-reduced EGB equations never lose hyperbolicity in the exterior of UBSs.

In Section~\ref{sec:Nonlinear_evo} we present some results of non-linear numerical simulations of perturbed black strings and discussing some aspects of the non-linear dynamics that are complementary to \cite{Figueras:2026epx}. In particular, we focus on the breakdown of hyperbolicity that terminates Cauchy evolution. For either sign of the Gauss-Bonnet coupling, the breakdown occurs inside the apparent horizon where the thinnest string segments join the satellite black holes, but the type of breakdown depends on the sign of the coupling. For $\lambda^\text{GB}<0$ the transition is of Keldysh type, whereas for $\lambda^\text{GB}>0$ the transition is of Tricomi type. In the positive coupling case the thinnest segments of the apparent horizon have a local coupling $\lambda^\text{GB}/r_H^2\approx0.08$ at the time of the breakdown, i.e., slightly below the threshold value $\tfrac14(\sqrt2-1)\approx0.104$ obtained from UBS backgrounds. We discuss the implications of our results and future directions in Section~\ref{sec:discussion}.

The rest of this article is organized as follows. In Section \ref{sec:UBS_EGB} we describe the general properties of UBS in EGB gravity, and in Section \ref{sec:Linear_analysis} we study their linear stability. Section \ref{sec:mccz4_cartoon} is devoted to the well-posedness of the cartoon equations. Some technical details have been relegated to the appendices; in Appendix \ref{app:gauge_fixing} we show that the spherical gauge employed in the study of the time-dependent perturbations can always be achieved and in Appendix \ref{app:evolution_equations} we list the symmetry-reduced evolution equations.
We use Greek letters $\alpha,\beta,\ldots$ to denote spacetime coordinates. The first letters of the Latin alphabet, $a,\,b\,,\ldots$ are used to denote the directions transverse to the round $S^2$ of the black strings. We use the curvature conventions of \cite{Wald:1984rg}.
 
\section{Uniform Black Strings in Einstein-Gauss-Bonnet gravity}
\label{sec:UBS_EGB}

In this section we outline the numerical construction of uniform black strings in EGB  gravity. Unlike in Einstein gravity (i.e., $\lambda^\text{GB}=0$), adding a flat direction to the four-dimensional Schwarzschild solution does not produce a solution to \eqref{eq:eomsEGB}. Therefore, to construct uniform black string solutions in EGB gravity we have to solve \eqref{eq:eomsEGB} numerically. To do so, we will employ the deTurck trick and add a gauge fixing term to \eqref{eq:eomsEGB}:
\begin{equation}
    R_{\mu\nu} -\nabla_{(\mu}\xi_{\nu)} + \lambda^\text{GB}\,H_{\mu\nu}=0\,, \label{eq:deTurckEGB}
\end{equation}
where $\xi^\mu = g^{\rho\sigma}(\Gamma^{\mu}_{\phantom{\mu}\rho\sigma}-\bar\Gamma^\mu_{\phantom{\mu}\rho\sigma})$, and $\Gamma^{\mu}_{\phantom{\mu}\rho\sigma}$ are the Christoffel symbols associated to the spacetime metric that we are trying to find, $g_{\mu\nu}$, and $\bar\Gamma^{\mu}_{\phantom{\mu}\rho\sigma}$ are the Christoffel symbols associated to a reference metric on the same spacetime manifold, $\bar g_{\mu\nu}$. In our calculations we take the uniform  Schwarzschild black string in five spacetime dimensions as the reference metric. The advantage of solving \eqref{eq:deTurckEGB} instead of \eqref{eq:eomsEGB} is that, in the case of stationary spacetimes, the former are expected to be an elliptic set of equations, at least for small enough values of the coupling $\lambda^\text{GB}$. Diffeomorphism invariance implies that, on any solution of \eqref{eq:deTurckEGB}, the vector $\xi^\mu$ satisfies
\begin{equation}
    \nabla^\nu\nabla_\nu\,\xi_\mu+R_{\mu\nu}\,\xi^\nu=0\,.
    \label{eq:xi_identity}
\end{equation}
With our boundary conditions, the operator in \eqref{eq:xi_identity} has a trivial kernel on the GR UBS. As we shall explain, this property is stable under small deformations of the background, implying that solutions of \eqref{eq:deTurckEGB} that are sufficiently close to the GR UBS have $\xi^\mu=0$ and are solutions of \eqref{eq:eomsEGB}. To see this, consider the class of static, $x$-independent perturbations. The linearisation of \eqref{eq:deTurckEGB} around the GR UBS is then the Lichnerowicz operator of the Euclidean UBS which has trivial kernel. Its only non-positive mode, inherited from the Euclidean Schwarzschild black hole \cite{Gross:1982cv}, has a non-zero eigenvalue, while the zero mode associated with the GL threshold has $k\neq0$. The implicit function theorem then implies that for sufficiently small $\lambda^\text{GB}/r_H^2$ there is a unique solution of \eqref{eq:deTurckEGB} in a neighbourhood of the GR UBS, and it is a UBS of EGB gravity.
Note that this continuity argument and the maximum-principle arguments of \cite{Figueras:2011va,Figueras:2016nmo} do not apply directly. However, the absence of DeTurck solitons persist as long as the corresponding linearized operators remain invertible which can be monitored numerically.

UBSs in five-dimensional EGB were first constructed in \cite{Kobayashi:2004hq}, see also \cite{Suranyi:2008wc,Brihaye:2010me}. They found that regular solutions only exist for a range of couplings $\lambda^\text{GB}$, see eq. \eqref{eq:range_UBS};

we verify these bounds in our construction, see below in Fig. \ref{fig:physQ_UBS}.  We have constructed static uniform black string solutions in EGB using  a metric ansatz of the form:
\begin{equation}
\begin{aligned}
    ds^2=&-r_0^2\,r^2\,A\,dt^2+\frac{4\,r_0^2\,B}{(1-r^2)^4}\,dr^2+\frac{r_0^2\,S}{(1-r^2)^2}\,d\Omega_{(2)}^2\\
    &+W\,dx^2\,, 
\end{aligned}
\label{eq:Schw_coords}
\end{equation}
where we have compactified the radial coordinate, $r\to \frac{r_0}{1-r^2}$, and $X\equiv\{A,\,B,\,S,\,W\}$ are the unknown functions, which only depend on the radial coordinate $r$. The coordinate $x$ parametrises the (periodic) direction along the string; asymptotically, we fix $x\sim x+L$, so $L$ is the proper length of the asymptotic Kaluza-Klein circle. The Schwarzschild black string in GR corresponds to $A=B=S=W=1$. In the coordinates \eqref{eq:Schw_coords}, the horizon of the black string is at $r=0$ and the functions $X$ should be regular there, i.e., $\partial_r X|_{r=0}=0$, together with $(A-B)|_{r=0}$ to avoid conical singularities. At infinity, $r=1$, we have $X=1$ to guarantee asymptotic flatness. 

\subsection{Physical quantities}
For general asymptotically Kaluza-Klein $D$-dimensional spacetimes,  the spacetime metric approaches $\text{Mink}_{D-1}\times S^1$ near infinity. From the fall offs of the $tt$ and $xx$ components of metric,
\begin{equation}
    g_{tt}\simeq -1+\frac{c_M}{\varrho^{D-4}}\,,\qquad g_{xx}\simeq 1-\frac{c_\tau}{\varrho^{D-4}}\,,
\end{equation}
where $\varrho$ is the radial coordinate in the $(D-1)$-dimensional Minkowski space, one can read off the conserved charges, namely mass $M$ and tension $\mathcal{T}$ \cite{Harmark:2003dg}:
\begin{equation}
\begin{aligned}
    M=&~\frac{\Omega_{D-3}\,L}{16\pi\,G}\big((D-3)\,c_M+ c_\tau \big)\,,\\
    \mathcal{T}=&~\frac{\Omega_{D-3}}{16\pi\,G}\big(c_M+(D-3)\,c_\tau \big)\,,
\end{aligned}
\end{equation}
where $\Omega_{D-3}$ is the area of the $(D-3)$-dimensional unit sphere. It is also useful to consider the relative tension $n$, 
\begin{equation}
    n = \frac{\mathcal{T}\,L}{M} = \frac{c_M+(D-3)c_\tau}{(D-3)c_M+c_\tau}\,.
\end{equation}
In terms of our metric anstaz \eqref{eq:Schw_coords} and for $D=5$, we can compute $c_M$ and $c_\tau$ as
\begin{equation}
    c_M=r_0\left(1+\tfrac{1}{2}\,\partial_r A|_{r=1}\right)\,,\quad c_\tau = \tfrac{r_0}{2}\,\partial_r W|_{r=1}\,.
\end{equation}
For the UBS in GR, $W(r)=1$, so $c_\tau=0$ and hence $n=\frac{1}{2}$. 

The inverse temperature of the black strings is given by
\begin{equation}
    \beta = 4\pi\,r_0\,.
\end{equation}
Note that in GR, because there is no intrinsic length scale, one can set $r_0=1$ without loss of generality. In EGB, however, there is a natural scale set by the choice of coupling $\lambda^\text{GB}$. Therefore, there exists a one parameter family of UBS labelled by the dimensionless number $\lambda^\text{GB}/r_0^2$. In practice, to construct the family of UBSs, we set $r_0=1$ and vary $\lambda^\text{GB}$.

The entropy of the UBS can be calculated using Wald's formula \cite{Wald:1993nt,Visser:1993nu,Jacobson:1993vj,Iyer:1994ys},
\begin{equation}
    S=-\frac{2\pi}{G}\int_H \left(\frac{\partial\mathscr{L}}{\partial R_{\mu\nu\rho\sigma}}\right)\epsilon_{\mu\nu}\,\epsilon_{\rho\sigma}\,,
\end{equation}
where $\epsilon_{\mu\nu}$ denotes the binormal to the horizon, i.e., the natural volume element of the tangent space orthogonal to the horizon. For our metric ansatz \eqref{eq:Schw_coords}, this formula evaluates to
\begin{equation}
    S=~\frac{A_H}{4\,G}\left(1+\tfrac{4\,\lambda^\text{GB}}{r_H^2}\right)\,,
\end{equation}
where $A_H$ is the area of the horizon and $r_H^2\equiv r_0^2\,S(0)$ is the radius (squared) of the horizon $S^2$. 

To characterize the black strings, we will also compute the Kretschmann and the cubic invariant, using the same normalization as in \cite{Lehner:2010pn}: 
\begin{align}
    \bar{\mathcal{K}}=&~\tfrac{1}{12}\,I\,r_H^4\,, \label{eq:norm_K} \\
    \bar{\mathcal{S}}=&~324\,J^2\,I^{-3} - 26\,, \label{eq:norm_S}
\end{align}
where $I\equiv R_{\mu\nu\rho\sigma}\,R^{\mu\nu\rho\sigma}$ is the Kretschmann invariant and $J\equiv R^{\mu\nu}{}_{\rho\sigma}\,R^{\rho\sigma}{}_{\alpha\beta}\,R^{\alpha\beta}{}_{\mu\nu}$. The normalizations of $\bar{\mathcal{K}}$ and $\bar{\mathcal{S}}$ have been chosen so that both of these quantities are equal to one for a UBS in GR.  As we shall see below,  $r_H$ is also a geometric quantity for the class of dynamical black strings that we consider in this article. 

\begin{figure}[t]
    \centering
    \includegraphics[width=0.9\linewidth]{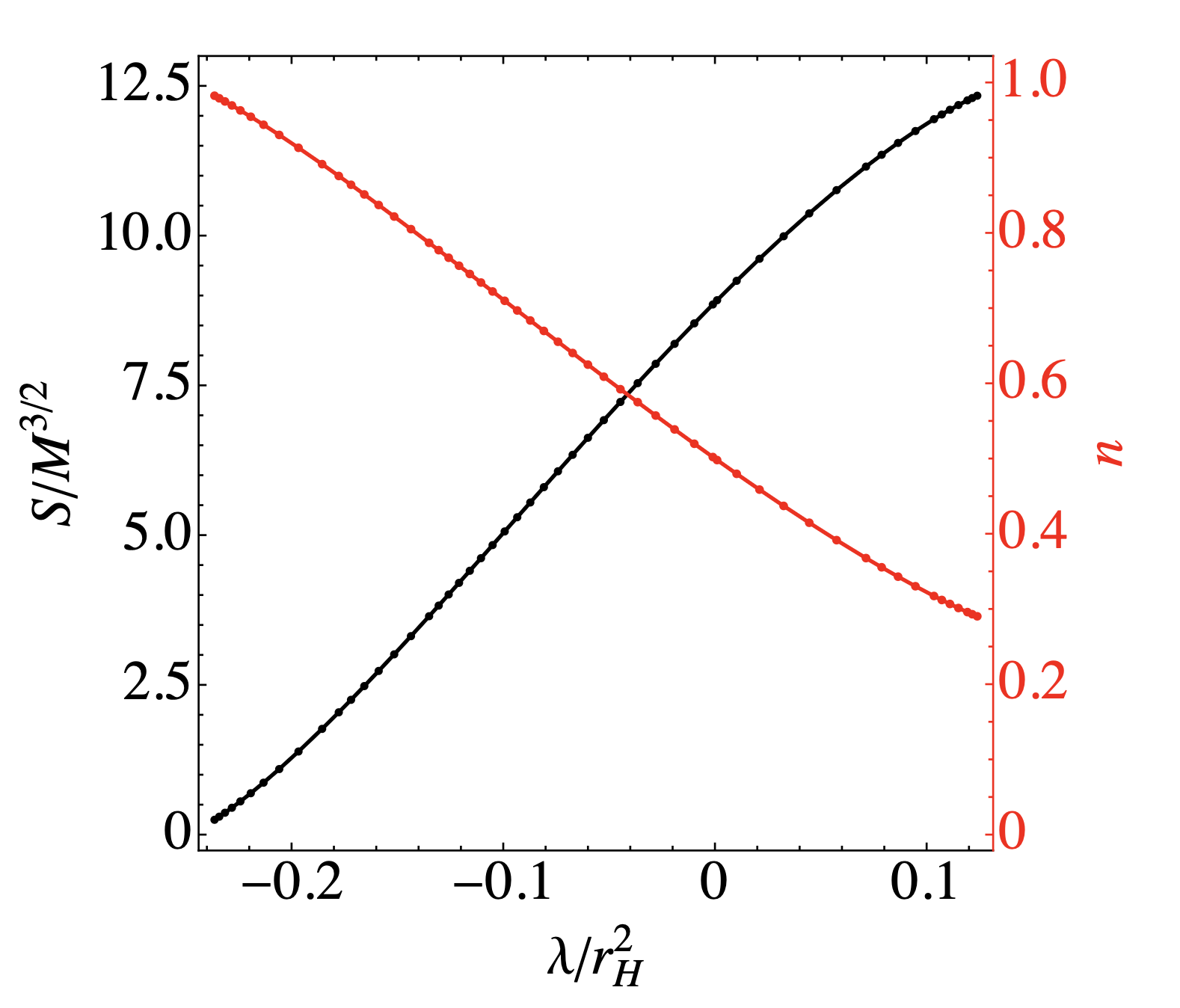}
    \includegraphics[width=0.9\linewidth]{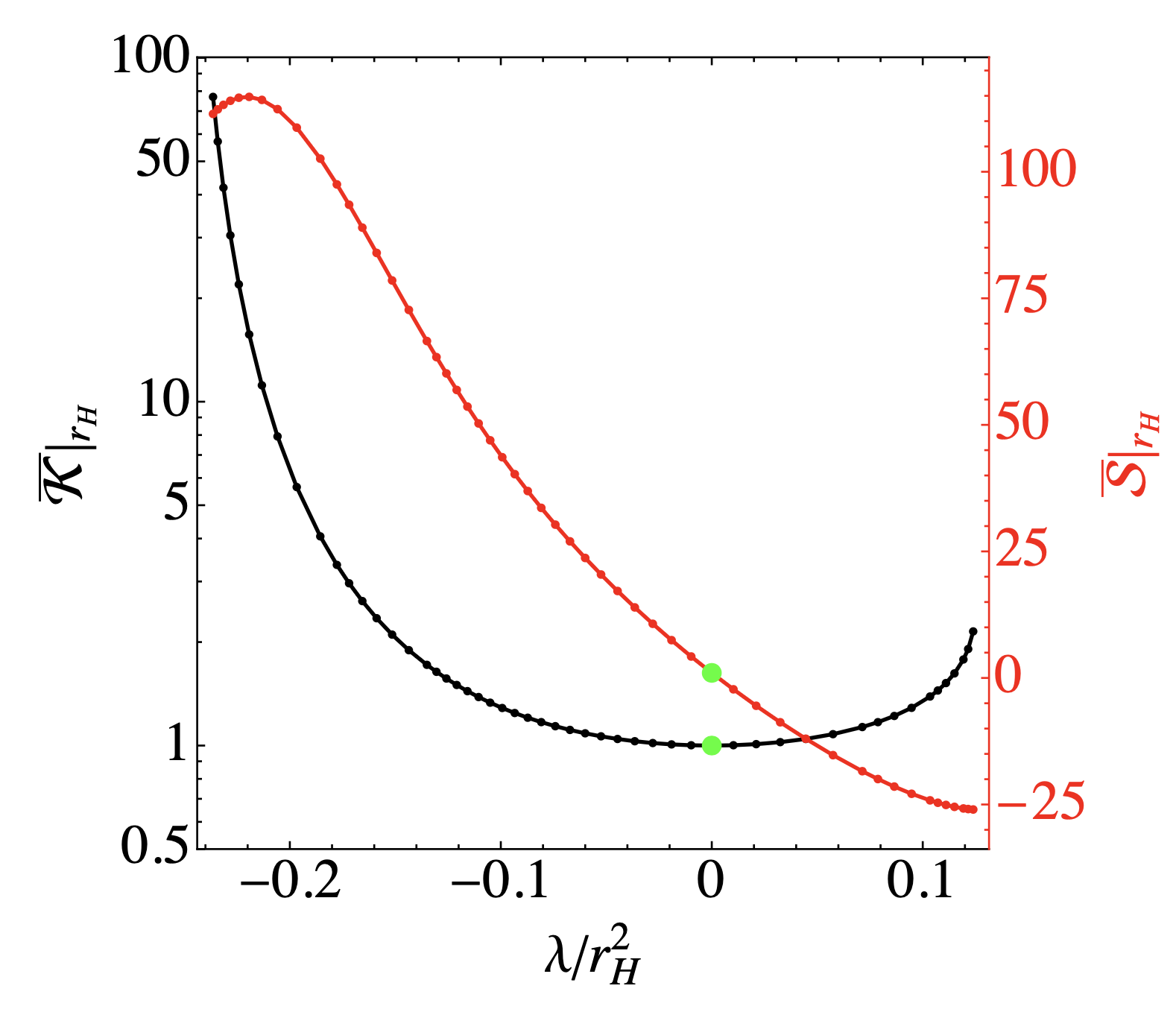}
    \caption{\textit{Top}: Dimensionless entropy $S/M^{3/2}$ and relative tension $n$ as functions of the coupling for UBS in EGB in five dimensions. For $\lambda^\text{GB}>0$, the solution at the end of the branch has finite entropy. For $\lambda^\text{GB}>0$, the string tension decreases while for $\lambda^\text{GB}>0$ it increases. \textit{Bottom}: normalized curvature invariants evaluated on the horizon as functions of the coupling. The green dot shows the UBS in GR. The curvature invariants diverge as $\lambda^\text{GB}/r_H^2\to -1/4$ while they remain finite in the limit $\lambda^\text{GB}/r_H^2\to 1/8$.}
    \label{fig:physQ_UBS}
\end{figure}

In Fig. \ref{fig:physQ_UBS} we plot the (dimensionless) entropy, $S/M^{3/2}$, and relative tension $n$ for UBS in EGB for different couplings.  This plot shows that the limiting solution $\lambda^\text{GB}/r_H^2\to 1/8$ has finite entropy and finite curvature invariants, whilst at the other end of the range, $\lambda^\text{GB}/r_H^2\to -1/4$, the solution is a naked singularity, with diverging curvature invariants.\footnote{Note that the regularity of the limiting solution at $\lambda^\text{GB}/r_H^2=1/8$ alone does not imply that it is relevant for the dynamics of black strings in EGB gravity. Static solutions are constructed by solving a boundary value problem which is insensitive to whether the time-dependent equations are hyperbolic around them or whether these solutions are dynamically attainable. We will return to this question in Section~\ref{subsec:hypberb}. } It is also interesting to note that the tension of strings decreases as the Gauss-Bonnet coupling $\lambda^\text{GB}$, while it increases for negative values of $\lambda^\text{GB}$. As we shall see below, this behavior is correlated with the stability properties of UBSs in EGB, and the subsequent non-linear evolution of these instabilities \cite{Figueras:2026epx}.

\section{Linear stability of UBS in EGB}
\label{sec:Linear_analysis}

In this section we study the linear (in)stability properties of UBS in EGB for different couplings, seeking appropriate mode solutions of the linearized equations around UBS backgrounds. First we consider time-independent perturbations (zero modes) and then we compute the growth rates of the time-dependent perturbations. As in GR, the zero modes signal the existence of a new branch of static non-uniform black strings \cite{Brihaye:2010me}. We note, however, that such solutions are only meaningful as approximations to solutions of the non-linear theory if the linearized initial value problem is well-posed. In EGB theory this is guaranteed only in a weakly coupled regime. As we shall see in Section~\ref{subsec:hypberb} (under spherical symmetry in the asymptotic directions) this condition fails near the horizon of the UBSs with $\lambda^\text{GB}/r_H^2>\tfrac14(\sqrt2-1)\approx 0.104$. Therefore, the results of this section for such large couplings should be regarded as formal solutions of the linearized equations with no direct dynamical meaning.

\subsection{Zero modes}

The zero modes of perturbed UBS in EGB were studied in \cite{Suranyi:2008wc}. This reference used the following ansatz for the background UBS: 
\begin{equation}
\begin{aligned}
    ds^2=&-r_0^2\,r^2\,f(r)\,dt^2+\frac{4\,r_0^2\,dr^2}{(1-r^2)^4 g(r)}+\frac{r_0^2}{(1-r^2)^2}\,d\Omega_{(2)}^2\\
    &+h(r)\,dx^2\,,
\end{aligned}
\label{eq:ansatz_UBS_2}
\end{equation}
where we have compactified the radial coordinate as before and the functions $f(r)$, $g(r)$ and $h(r)$ have to be determined numerically by solving the equations of  motion \eqref{eq:eomsEGB}. Note that in this ansatz the parameter $r_0$ corresponds to the radius of the horizon $S^2$, i.e., $r_H$, but the temperature measured in asymptotically normalised time is $T=\sqrt{f(0)g(0)/f(1)}/(4\pi r_0)$. 

We consider the following (partially gauge-fixed) form of the static perturbations of \eqref{eq:ansatz_UBS_2}:
\begin{equation}
    \begin{aligned}
        ds^2=&-r_0^2\,r^2\,f(r)\,e^{2A}dt^2+\frac{4\,r_0^2\,e^{2B}}{(1-r^2)^4 g(r)}\,dr^2\\
        &+\frac{r_0^2\,e^{2S}}{(1-r^2)^2}\,d\Omega_{(2)}^2+h(r)\,e^{2W}\,dx^2\,,
    \end{aligned}
    \label{eq:anstaz_NUBS}
\end{equation}
where 
\begin{equation}
\begin{aligned}
    A=&~\epsilon\,a_1(r)\,\cos(k\,x)\,,\\
    B=&~\epsilon\,b_1(r)\,\cos(k\,x)\,,\\
    S=&~\epsilon\,s_1(r)\,\cos(k\,x)\,,\\
    W=&~\epsilon\,w_1(r)\,\cos(k\,x)\,,
\end{aligned}
\label{eq:static_linear_h}
\end{equation}
and $\epsilon$ is a small parameter. Here $a_1(r)$, $b_1(r)$, $s_1(r)$ and $w_1(r)$ are the unknown functions that parametrize the linear metric perturbations. 

We plug the ansatz \eqref{eq:anstaz_NUBS} into the equations of motion \eqref{eq:eomsEGB} and expand the latter to linear order in $\epsilon$ using \eqref{eq:static_linear_h}. To fully fix the gauge, we impose
\begin{equation}
    s_1(r)=-\tfrac{1}{2}\,a_1(r)\,.
\end{equation}
This corresponds to the so called Kol-Sorkin gauge fixing \cite{Kol:2006vu}. The main advantage of this gauge is that one can obtain a single master equation for $a_1(r)$ that governs the perturbation. Furthermore, this equation only has two singular points, namely the horizon $r=0$ and infinity $r=1$. In this gauge, regularity at the horizon simply requires $\partial_ra_1(r)|_{r=0}=0$, while the decaying solution at infinity behaves like
\begin{equation}
    a_1(r)\approx e^{-\alpha/(1-r) }(1-r)^\beta\sum_{n=0}^\infty \bar a_n(1-r)^n\,,
    \label{eq:a1_near_infty}
\end{equation}
with\footnote{Here we assume $k>0$ without loss of generality.}
\begin{equation}
    \begin{aligned}
        \alpha=&~\frac{k\,r_0}{2\,\bar h_0^\frac{1}{2}}\,,\\
        \beta =&~\tfrac{1}{4\,\bar f_0}\left[k\,r_0\,\bar f_1+\bar f_0\big( 2\,k\,r_0(1-\bar g_1)+4\,\bar h_0^\frac{1}{2}\big)\right]\,,
    \end{aligned}
\end{equation}
and where $\bar h_0$, $\bar f_0$, $\bar f_1$ and $\bar g_1$ are the coefficients in the expansions of the background metric functions in \eqref{eq:ansatz_UBS_2} near infinity,
\begin{equation}
    X(r)=\sum_{n=0}^\infty \bar x_n\,(1-r)^n\,,
\end{equation}
with $X(r)=\{f(r),\,g(r),\,h(r)\}$. Such coefficients  can be determined by solving the equations of motion \eqref{eq:eomsEGB} perturbatively near $r=1$.

Imposing regularity at the horizon and simple Dirichlet boundary conditions at infinity for $a_1(r)$, the master equation becomes an eigenvalue equation where the eigenvalue is $k^2$, and can be straighforwardly solved. Alternatively, one can build in the behavior of $a_1(r)$ near infinity, eq.\eqref{eq:a1_near_infty}, and obtain an equation for a regular function $q(r)$ defined as
\begin{equation}
    a_1(r)= e^{-2\alpha/(1-r^2)}(1-r^2)^{\beta-1}q(r)\,.
\end{equation}
Since $2/(1-r^2)=1/(1-r)+1/(1+r)$, this substitution removes the leading exponential in \eqref{eq:a1_near_infty} and $q(r)=O(1-r)$ at infinity. Thus $q(r)$ satisfies a Neumann boundary condition $q^{\prime}(0)=0$ at the horizon and a Dirichlet boundary condition $q(1)=0$ at infinity. The resulting equation for $q(r)$ can no longer be cast as an eigenvalue problem, but it can be straightforwardly solved with the Newton-Raphson algorithm, where the critical wavenumber $k_c$ is one of the unknowns. 

\begin{figure}[t]
    \centering
    \includegraphics[width=0.9\linewidth]{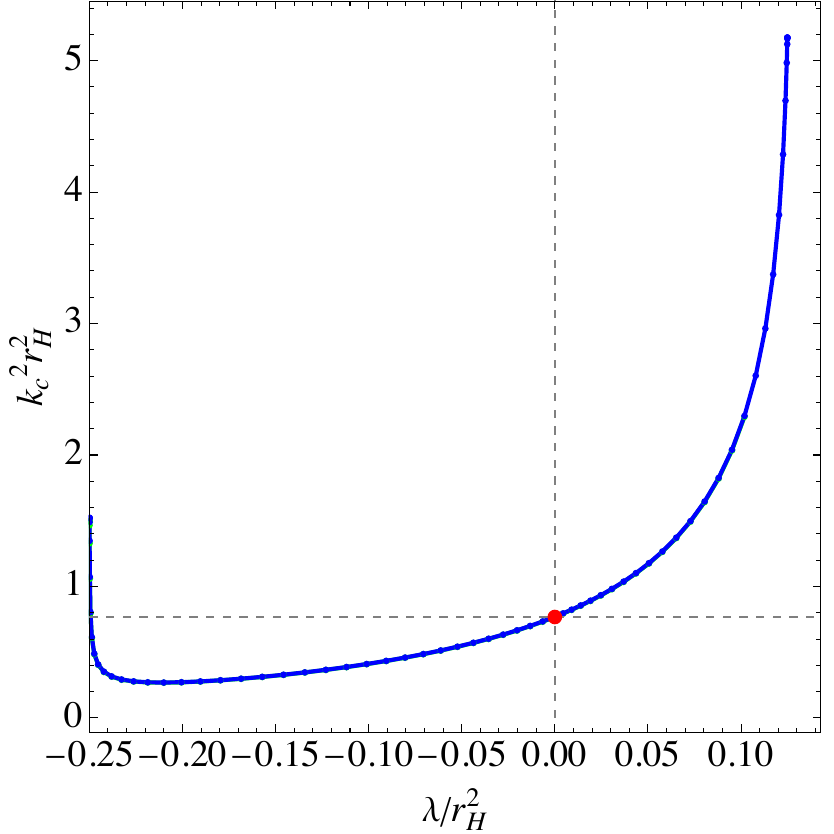}
    \caption{Critical wave number $k_c^2\,r_H^2$ as a function of the Gauss-Bonnet coupling constant $\lambda$ obtained by solving the eigenvalue problem or the equation for a regular function with the Newton-Raphson method. The results obtained with both methods overlap. The critical wave number sharply increases near both limiting values of the coupling. }
    \label{fig:zero_modes}
\end{figure}

We have implemented both methods and the results that we obtain for the zero mode $k_c$ agree with each other and with those first presented in \cite{Suranyi:2008wc}, see Fig. \ref{fig:zero_modes}. We have also computed the critical value $k_c$ by taking the limit $\Omega\to 0$ of the time-dependent perturbations (see the next subsection), where $\Omega$ is the growth rate of the unstable modes, and all these different ways of computing $k_c$ produce the same results.  

The results in Fig. \ref{fig:zero_modes} show that $k_c$ increases monotonically for larger and positive values of the $\lambda^\text{GB}$, and sharply increases as the limiting value of $\lambda^\text{GB} = \frac{1}{8}r_H^2$ is reached. This behavior of $k_c$ indicates that shorter wavelengths should become unstable; this is intuitive since these strings have lower tension and hence they are ``easier'' to deform. On the other hand, $k_c$ decreases as $\lambda^\text{GB}$ decreases, become more negative, so only longer wavelengths should be unstable and hence thinner UBS are stable; again, this behvior is intuitive since now these strings have larger tension. This behavior of the zero mode for $\lambda^\text{GB}<0$ changes rather abruptly near $\lambda^\text{GB}\approx -0.23 r_H^2$; from this point and for more negative values of the coupling,  $k_c$ starts increasing again and the slope of the curve appears to diverge as $\lambda^\text{GB} \to -\frac{1}{4}r_H^2$, indicating that the UBSs are becoming more unstable again, i.e., shorter wavelengths are unstable. This can be understood because the background UBS is becoming singular in this limit. 

\subsection{Time-dependent perturbations}\label{sec:UBS_GL}

In this subsection we study the time-dependent perturbations of UBS in EGB, extending the perturbative computations of \cite{Henriquez-Baez:2022bfi}. The growth rates of the GL instability that we obtain in this subsection (as functions of the coupling) are useful to  interpret the results of the fully non-linear evolution of these instabilities presented in \cite{Figueras:2026epx}. 

To study the time-dependent perturbations, we consider the background UBS in the coordinates  \eqref{eq:Schw_coords} and we impose the spherical gauge for the metric perturbations $h_{\mu\nu}$ \cite{Hovdebo:2006jy}:
\begin{equation}
    \begin{aligned}
    h_{tt}=&~e^{\Omega\,t}\,\cos(k\,x)\,H_t(r)\,,\\
    h_{tr}=&~e^{\Omega\,t}\,\cos(k\,x)\,\Omega\,H_v(r)\,,\\
    h_{rr}=&~e^{\Omega\,t}\,\cos(k\,x)\,H_r(r)\,,\\
    h_{rx}=&~e^{\Omega\,t}\,\sin(k\,x)\,k\,H_v(r)\,,\\
    h_{xx}=&~e^{\Omega\,t}\,\cos(k\,x)\,H_x(r)\,,
    \end{aligned}
    \label{eq:spherical_gauge}
\end{equation}
and all the other components are zero. In Appendix \ref{app:gauge_fixing} we show how the general metric perturbations can be brought into this particular gauge. 

Whilst it is possible to derive,  from the linearized equations of motion, a single master equation for $H_x(r)$ that governs the metric perturbations, the actual expression is very cumbersome and exceedingly long (due to the contributions from the Gauss-Bonnet terms). Here we follow a variant of the  approach proposed in \cite{Chung:2023wkd}. The idea is the following. Having fully fixed the gauge, one then solves \textit{all} components of the linearized equations of motion at once, thus avoiding having to find a decoupled equation for a single variable that governs all the metric perturbations. The system is over determined since there are more equations than variables and some of the equations are only first order in derivatives (they are constraints), which implies that the linear operator acting on the metric perturbations is a rectangular matrix.  One can simultaneously solve all the components of the linearized equations of motion in terms of all components of the metric perturbations using a Newton-Raphson algorithm by inverting the (discretised) linear operator at every step with the Moore-Penrose inverse. This method is particularly useful when computing perturbations of black holes in effective field theories of gravity since the linearized equations of motion tend to be significantly more complicated than in general relativity and deriving a single master equation for a master variable may not be possible in general. The effectiveness of this method was demonstrated in \cite{Chung:2023wkd}, and subsequently in \cite{Chung:2024vaf,Chung:2025gyg}, where they computed the quasinormal modes of four-dimensional rotating black holes in Einstein-scalar-Gauss-Bonnet gravity and in Dynamical Chern-Simons gravity respectively. The only (minor) difference between their approach and ours is that in these references they explicitly expand the metric perturbations in a basis of orthogonal polynomials and obtain the coefficients in the expansion by solving the linearized equations of motion. In our case, we solve the linearized equations of motion at the collocation points. The two approaches are entirely equivalent and they are simply related by a (discrete) Fourier transform. 

Imposing regularity at the horizon and the correct asymptotic behavior at infinity imply that the metric perturbations should behave as
\begin{equation}
    \begin{aligned}
        H_x(r)=&~r^{2\,\Omega}\,e^{-\sqrt{k^2+\Omega^2}/(1-r^2)}(1-r^2)^\beta\,Q_x(r)\,,\\
        H_t(r)=&~r^{2\,\Omega}\,e^{-\sqrt{k^2+\Omega^2}/(1-r^2)}(1-r^2)^\beta\,Q_t(r)\,,\\
        H_r(r)=&~r^{2\,\Omega-2}\,e^{-\sqrt{k^2+\Omega^2}/(1-r^2)}(1-r^2)^{\beta-2}\,Q_r(r)\,,\\
        H_v(r)=&~r^{2\,\Omega-1}\,e^{-\sqrt{k^2+\Omega^2}/(1-r^2)}(1-r^2)^{\beta-2}\,Q_v(r)\,,
    \end{aligned}
\end{equation}
where the $Q$'s are regular functions at both the horizon ($r=0$) and infinity ($r=1$), and
\begin{equation}
    \begin{aligned}
    \beta=&~\tfrac{1}{24\sqrt{k^2+\Omega^2}}\Bigl[ 
    4\,k^2+8\bigl(\Omega^2-6\sqrt{k^2+\Omega^2}\bigr)+9\,k^2\,W_1\\
    &\hspace{1.75cm}+(k^2+2\,\Omega^2)\sqrt{64+9\,W_1^2}
    \Bigr]\,,
    \end{aligned}
\end{equation}
with $W_1\equiv \partial_r W|_{r=1}$.

\begin{figure}[t]
    \centering
    \includegraphics[width=0.98\linewidth]{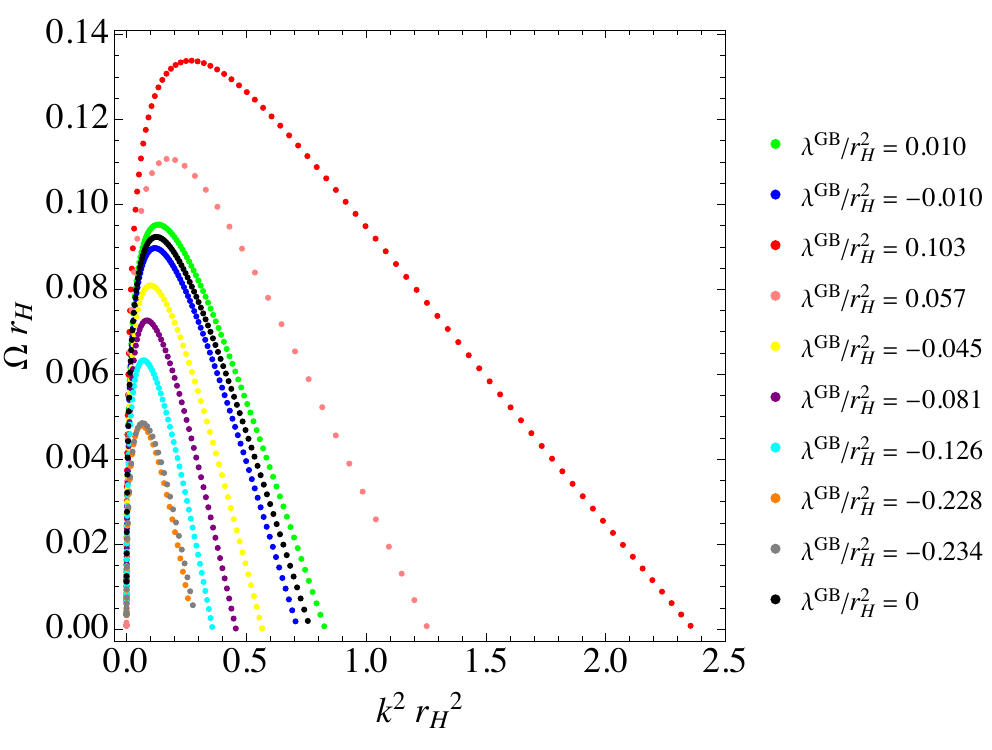}
    \caption{Growth rates of the GL instabilities of UBSs in EGB gravity for different values of the coupling. For positive couplings, the growth rates become larger and higher wavenumbers $k$ (i.e., shorter wavelengths) become unstable. For negative couplings, the opposite happens, and the growth rates become smaller and the instability shifts towards longer wavelengths. However, this trend is reversed for a sufficiently negative coupling, and the growth rates increase again and shorter wavelengths become unstable as the background approaches a singular limit. Results for positive couplings are shown up to the hyperbolicity threshold $\lambda^\text{GB}/r_H^2<\tfrac14(\sqrt2-1)\approx0.104$ (see Section~\ref{subsec:hypberb}).}
    \label{fig:growth_rates}
\end{figure}

To proceed, we use a Newton-Raphson algorithm to numerically solve all the linearized equations of motion for the $Q$'s using standard pseudospectral methods. We impose the boundary condition $Q_x|_{r=1}=1$, which then allows us to treat one of the parameters, namely $k$ or $\Omega$, as an unknown while we vary the other one. The results for some representative values of the Gauss-Bonnet coupling $\lambda^\text{GB}$ are shown in Fig. \ref{fig:growth_rates}. We restrict to couplings $\lambda^\text{GB}/r_H^2<\tfrac14(\sqrt2-1)$ since the linearized initial value problem around the UBS is ill-posed for larger values of the coupling (see Section~\ref{subsec:hypberb}). Above this bound growth rates can still be computed but the corresponding mode solutions do not admit an interpretation of describing an approximate non-linear solution arising from generic nearby data.

Figure \ref{fig:growth_rates} confirms the expectations that we drew from the study of the zero modes. As this figure shows, the UBS are indeed more unstable, i.e., the instability is faster and shorter wavelengths become unstable, the larger the  positive values of the coupling $\lambda^\text{GB}$ are. This can be understood from the fact that the tension of the strings monotonically decreases for $\lambda^\text{GB}>0$, which makes it easier for the string to be deformed dynamically, and become longer (and thinner) under evolution. This expectation is confirmed in the non-linear evolution of the instability \cite{Figueras:2026epx}. On the other hand, for $\lambda^\text{GB}<0$, the instability is weaker, i.e., the growth rates decrease and wavelengths of the unstable modes increase. This is consistent with the fact that the tension for such strings increases as the coupling becomes more negative, which implies that it is harder to deform them. This behaviour changes for values the coupling close $-\frac{1}{4}r_H^2$, where one would expect that the strings should become more unstable again because the critical wavenumber $k_c$ sharply increases. In this case the reason for this is that the background in UBS is becoming singular as one approaches the minimum allowed value of $\lambda^\text{GB}$.

\section{Well-posedness of Black String Evolution in EGB}
\label{sec:mccz4_cartoon}

The linear analysis of Sec.~\ref{sec:Linear_analysis} provides a picture of the 
GL instability of uniform black strings in the presence of the Gauss-Bonnet correction. 
To investigate the subsequent non-linear dynamics, it is crucial that the theory admits a well-posed initial value formulation, which in turn demands that the evolution system be strongly hyperbolic. In this section, we revisit the mCCZ4 formulation of EGB gravity \cite{AresteSalo:2022hua,AresteSalo:2023mmd} and analyse its hyperbolicity in spacetimes with spherical symmetry in the asymptotically flat directions. This framework provides the basis for our non-linear simulations in \cite{Figueras:2026epx} and enables us to diagnose the dynamical loss of hyperbolicity observed during the evolution. The general case, without the symmetry reduction, is discussed in \cite{SY_paper}.

\subsection{mCCZ4 formulation of EGB Gravity} 
\label{subsec:formulation}
The mCCZ4 formulation of the Einstein equations \cite{AresteSalo:2022hua,AresteSalo:2023mmd} is a generalisation to singularity avoiding coordinates of the modified harmonic formulation introduced in \cite{Kovacs:2020pns,Kovacs:2020ywu}. Using these formulations, it can be shown that certain deformations of GR, such as  Horndeski scalar-tensor theories of gravity or EGB, have a well-posed initial value problem in the weakly coupled regime. In mCCZ4,  the covariant field equations \eqref{eq:eomsEGB} in $d+1$ dimensions are extended by Z4 terms and supplemented by gauge conditions for the lapse $\alpha$ and shift $\beta^i$,\footnote{Here we choose the so called ``integrated moving punctures'' gauge.} yielding the full system
\begin{subequations}
\begin{gather}
R^{\mu\nu}-\tfrac{1}{2}R\,g^{\mu\nu} +\lambda^{\text{GB}}\bar H^{\mu\nu}
+2\hat{P}_{\alpha}^{~\beta\mu\nu}\nabla_{\beta}Z^{\alpha}\nonumber\\
-\kappa_1\Big[2\,n^{(\mu}Z^{\nu)}+\Big(\tfrac{d-3}{2+b(x)}+\tfrac{d-1}{2}\kappa_2\Big)n^{\alpha}Z_{\alpha}\,g^{\mu\nu}\Big]
=0 ,
\label{eq:eoms_lovelock}
\\
\partial_{\perp}\alpha=\tfrac{c_{\alpha}}{1+a(x)}\alpha(K-2\Theta), \label{eq:eoms_gauge1} \\ 
\partial_{\perp}\beta^i=\tfrac{c_{\beta}}{1+a(x)}\,\hat{\Gamma}^i-\tfrac{a(x)}{1+a(x)}\,\alpha\,D^i\alpha+\eta\,\beta^i \,,\label{eq:eoms_gauge2}
\end{gather}
\end{subequations}
where $\bar H_{\mu\nu}=H_{\mu\nu}-\frac{1}{2}H g_{\mu\nu}$, $H_{\mu\nu}$ is given in \eqref{eq:def_H}, $\kappa_1$ and $\kappa_2$ are damping parameters, and $c_\alpha$ and $c_\beta$ determine the gauge. Here, $Z^{\mu}$ denotes the Z4 constraint vector, $n^{\mu}=\alpha^{-1}(\delta_t^{\mu}-\beta^i\delta_i^{\mu})$ is the unit vector normal to the hypersurfaces $t\equiv x^0=\text{const.}$, and $\partial_{\perp}=\partial_t-\beta^i\partial_i$ is the normal derivative to these hypersurfaces.
The scalar $\Theta:=-n_{\mu}Z^{\mu}$ and the spatial vector
$\hat{\Gamma}^i:=\tilde{\Gamma}^i+2\tilde{\gamma}^{ij}Z_j$,
where $\tilde{\Gamma}^i$ are the contracted Christoffel symbols of the conformal spatial metric $\tilde{\gamma}_{ij}$, encode the normal and spatial components of the Z4 constraint vector that enter the evolution equations.
The tensor 
\begin{equation}
    \hat{P}_{\alpha}^{~\beta\mu\nu}=\delta_{\alpha}^{(\mu}\,\hat{g}^{\nu)\beta}-\tfrac{1}{2}\,\delta_{\alpha}^{\beta}\,\hat{g}^{\mu\nu}
\end{equation} 
is defined in terms of the auxiliary metric 
\begin{equation}
\hat{g}^{\mu\nu}:=g^{\mu\nu}-b(x)\,n^{\mu}n^{\nu}\,, \label{eq:auxiliary_metric1}
\end{equation}
which governs the propagation of the gauge-condition-violating modes \cite{Kovacs:2020pns,Kovacs:2020ywu}. 
An auxiliary metric of the same form,
\begin{equation}
\tilde{g}^{\mu\nu}:=g^{\mu\nu}-a(x)\,n^\mu n^\nu\,,
\label{eq:auxiliary_metric2}
\end{equation}
was used in the modified harmonic formulation
\cite{Kovacs:2020pns,Kovacs:2020ywu} to define the \textit{modified harmonic gauge}
$\tilde{g}^{\alpha\beta}\Gamma^\mu_{\alpha\beta}=0$.
In the mCCZ4 formulation, $a(x)$ is introduced analogously to modify the
moving puncture gauge 
\eqref{eq:eoms_gauge1}--\eqref{eq:eoms_gauge2}. The system  is closed in the sense that the spatial and normal projections of \eqref{eq:eoms_lovelock} determine the evolution of the spatial metric and the Z4 constraint variables, while the gauge conditions \eqref{eq:eoms_gauge1}--\eqref{eq:eoms_gauge2} evolve the lapse and shift with the right-hand sides explicitly depending on $\Theta$ and $\hat{\Gamma}^i$, thus coupling the gauge dynamics to the Z4 sector.

Strong hyperbolicity requires the principal symbol $M(\xi_i)$ of a suitable first-order reduction to admit a smooth positive-definite symmetrizer, with upper and lower bounds uniform in unit spatial covectors and locally uniform in the background. Our companion work~\cite{SY_paper} establishes this property for mCCZ4 in weakly coupled EGB gravity by proving its equivalence, at the level of strong hyperbolicity, to a second-order metric system and adapting the analysis of \cite{Kovacs:2020ywu}. This completes the perturbative argument of \cite{AresteSalo:2023mmd}.

For the gauge conditions \eqref{eq:eoms_gauge1}--\eqref{eq:eoms_gauge2}, the characteristic analysis in \cite{SY_paper} separates the physical, pure-gauge, and gauge-condition-violating sectors. The temporal, longitudinal, and transverse pure-gauge modes satisfy $\mathcal{G}_{s}^{\mu\nu}\xi_\mu\xi_\nu=0$, with $s\in\{\bf 0,\bf{L},\bf{T}\}$ and auxiliary inverse metrics
\begin{equation}
\begin{aligned}
\mathcal{G}_{\bf 0}^{\mu\nu}
 &=\gamma^{\mu\nu}+\frac{(1+a(x))\alpha}{c_\alpha}n^\mu n^\nu,\\
\mathcal{G}_{ \bf L}^{\mu\nu}
 &=\gamma^{\mu\nu}-\frac{d(1+a(x))\alpha^2\chi}{2(d-1)c_\beta}n^\mu n^\nu,\\
\mathcal{G}_{\bf T}^{\mu\nu}
 &=\gamma^{\mu\nu}-\frac{(1+a(x))\alpha^2\chi}{c_\beta}n^\mu n^\nu.
\end{aligned}
\label{eq:pure_gauge_metrics}
\end{equation}
Here $\xi_\mu=(\xi_0,\xi_i)$ and $\chi$ is the conformal factor in $\tilde\gamma_{ij}=\chi\gamma_{ij}$. The metrics \eqref{eq:pure_gauge_metrics} are the puncture-gauge counterparts of \eqref{eq:auxiliary_metric2}: all of them can be written in the form $\gamma^{\mu\nu}-\epsilon_s n^\mu n^\nu$, with the coefficient $\epsilon_s$ determined by the corresponding gauge propagation speed. The gauge-condition-violating modes satisfy $\hat g^{\mu\nu}\xi_\mu\xi_\nu=0$, with $\hat g^{\mu\nu}$ given by \eqref{eq:auxiliary_metric1}. For fixed background and gauge parameters, these unphysical characteristic cones have no explicit dependence on $\lambda^{\mathrm{GB}}$.

We choose the gauge parameters so that the auxiliary metrics are Lorentzian, the time slices are non-characteristic with respect to them, and their characteristic cones are uniformly separated from one another and from the GR physical cone. Under these assumptions, \cite{SY_paper} constructs complete unphysical eigenspaces and a positive-definite Hermitian form on the physical total generalized eigenspaces, yielding a smooth symmetrizer at sufficiently weak coupling.

This argument establishes a weak-coupling neighbourhood of strong hyperbolicity without explicitly determining the physical characteristic roots. When symmetry simplifies the equations, these roots and their polarizations can be obtained directly using the gauge-invariant characteristic analysis of \cite{Reall:2021voz} (see also \cite{Kovacs:2020ywu,Thaalba:2024crk}). The physical polarizations satisfy
\begin{equation}
    \mathcal{P}^{\mu\nu\rho\sigma}(\xi)t_{\rho\sigma}\equiv P^{\mu\nu\alpha\beta\rho\sigma}\xi_{\alpha}\xi_{\beta} t_{\rho\sigma}=0,
    \label{eq:charaEq}
\end{equation}
where
\begin{equation}
    P^{\mu\nu\alpha\beta\rho\sigma}:=\frac{\partial E^{\mu\nu}}{\partial(\partial_\alpha\partial_\beta g_{\rho\sigma})},
\end{equation}
and $E^{\mu\nu}$ is the l.h.s. of the covariant equations of motion \eqref{eq:eomsEGB}. Here, $\mathcal{P}(\xi)^{\mu\nu\rho\sigma}$ is understood as an operator whose domain is the quotient of the space of symmetric rank-2 tensors by the subspace spanned by the ``high-frequency gauge transformations'' of the form $\xi_{(\rho}X_{\sigma)}$. Namely, eq.~\eqref{eq:charaEq} is defined for equivalence classes $[t_{\rho\sigma}]$, with the equivalence relation given by $t_{\rho\sigma}\sim t_{\rho\sigma}+\xi_{(\rho}X_{\sigma)}$.
A non-trivial solution has $[t_{\rho\sigma}]\neq0$. For a physical characteristic covector separated from the unphysical cones, this class admits a representative satisfying the principal mCCZ4 gauge condition. It then determines an eigenvector $(t_I,\xi_0t_I)^T$ of $M(\xi_i)$, where $t_I$ lists the components of that representative~\cite{SY_paper,Kovacs:2020ywu}.

As shown in the following subsection, $SO(3)$-invariant spacetimes on the $(2+1)$-dimensional reduced manifold have a single physical degree of freedom. Wherever the reduction is nonsingular, its characteristic equation is a quadratic form whose coefficients define an \textit{effective metric} $g_{\textbf{eff}}^{ab}$, in the sense of \cite{Reall:2021voz}. If the two roots $\xi_0^\pm$ are real and distinct for every nonzero spatial covector, the physical eigenspaces $V^\pm$ each have dimension one. Constructing a nonzero polarization for each root therefore diagonalizes the physical sector. A smooth polarization construction with uniformly separated roots for unit spatial covectors yields a smooth physical-sector symmetrizer with the bounds stated above. The effective metric thus provides an explicit physical-sector diagnostic on individual backgrounds, analogous to those used in scalar-tensor theories~\cite{Ripley:2019hxt,Ripley:2019irj,Bernard:2019fjb,Figueras:2020dzx}. Strong hyperbolicity of the full formulation additionally requires regular unphysical eigenspaces and separation from the physical spectrum. We now construct the polarization and derive this effective metric.

\subsection{Symmetry Reduction and Effective Metric}
In the study of the Gregory-Laflamme instability, we restrict attention to dynamics preserving the $SO(3)$ symmetry of the UBS. The corresponding spacetime metric can be written as
\begin{equation}
    ds^2 = g_{ab}dx^a dx^b + R^2 d\Omega^2,
    \label{eq:sym_reduced_metric}
\end{equation}
where $d\Omega^2$ is the metric on the round unit 2-sphere, $R(x^a)$ is the areal radius, and $x^a=(t,x,r)$ are coordinates on the symmetry-reduced manifold, with $x$ and $r$ denoting the periodic and radial directions of the black string respectively.

Although this form of the metric makes the symmetry explicit, it is more practical, for both numerical evolution and the hyperbolicity analysis, to  adopt Cartesian coordinates instead of the spherical-type chart $(t,x,r,\theta,\phi)$, as is standard in higher-dimensional numerical relativity \cite{Lehner:2010pn,Shibata:2010wz,Figueras:2015hkb,Cook:2016soy,Figueras:2017zwa,Bantilan:2019bvf,Andrade:2020dgc,Figueras:2022zkg}. To this end, we introduce coordinates $(z,w_1,w_2)$ through
\begin{equation}
    z=r\cos\theta \,,\quad
    w_1=r\sin\theta\cos\phi\,,\quad
    w_2=r\sin\theta\sin\phi,
\end{equation}
and implement the modified \textit{ cartoon reduction}  of \cite{Pretorius:2004jg,Yoshino:2009xp,Cook:2016soy} by restricting the system to the slice $w_1=w_2=0$, $z\ge0$. On this slice, $z = r$ serves as the radial coordinate.

Following \cite{Cook:2016soy}, the $SO(3)$ Killing symmetry implies that the only independent metric components are $g_{ab}$ and $g_{w_1w_1}=g_{w_2w_2}\equiv g_{ww}$, with $a,b\in\{t,x,z\}$, while the mixed components $g_{w_p b}$ vanish, with $p\in\{1,2\}$.
The same symmetry also determines all derivatives in the $w$-directions algebraically. In particular, $\partial_{w_p}g_{ab}=\partial_{w_p}g_{ww}=0$, and the only non-vanishing derivatives are\footnote{We stress that all of the equalities outlined here hold only on the $w_1 = w_2 = 0$ slice.} 
\begin{subequations}
    \begin{gather}
         \partial_{w_p} g_{w_q b}
       =\delta_{pq}\,\tfrac{g_{zb}-\delta^z{}_b\,g_{ww}}{z}, \\
       \partial_{w_p}\partial_{w_q} g_{ab}
       =\delta_{pq}\!\left(
          \tfrac{\partial_z g_{ab}}{z}
          -\tfrac{\delta^z{}_a\,g_{zb}+\delta^z{}_b\,g_{az}-2\delta^z{}_a\delta^z{}_b\,g_{ww}}{z^2}
       \right),\\
        \partial_{w_p}\partial_{w_q} g_{ww}
       =\delta_{pq}\!\left(
          \tfrac{\partial_z g_{ww}}{z}
          +2\,\tfrac{g_{zz}-g_{ww}}{z^2}
       \right),  
    \end{gather} \label{eq:ctrelation}
\end{subequations}
with $p,q\in\{1,2\}$. Thus, all $w$-derivatives appearing in the full equations of motion can be expressed in terms of lower order derivatives with respect to the $z$ coordinate, and hence the symmetry-reduced system is closed as a second-order PDE system for $(g_{ab},g_{ww})$. Accordingly, the principal part of the symmetry-reduced system inherits the structure of the original equations with the $w$-derivatives removed, because the latter are lower order terms as a consequence of the cartoon relations \eqref{eq:ctrelation}. 
Hence the resulting characteristic equation in the symmetry-reduced setting is just the restriction of the full characteristic equation, eq.\eqref{eq:charaEq}, to $SO(3)$-invariant polarizations, namely $(t_{ab},t_{ww})$, and the covector $\xi$ tangent to the reduced manifold, i.e., $\xi_\mu=(\xi_a,0,0)$.

 We now derive the reduction of \eqref{eq:charaEq} using the cartoon relations. As shown in \cite{Reall:2021voz,Kovacs:2020ywu}, the characteristic equation \eqref{eq:charaEq} can be written in the form 
\begin{equation}
    (\mathcal{P}(\xi)\cdot t )^{\mu}_\nu= \tfrac{1}{2}\bigl(\delta^{\mu\mu_1\mu_2}_{\nu\nu_1\nu_2}+\lambda^{\mathrm{GB}}\delta^{\mu\mu_1\mu_2\mu_3\mu_4}_{\nu\nu_1\nu_2\nu_3\nu_4} R^{\nu_3\nu_4}_{\mu_3\mu_4}\bigr)\xi^{\nu_1}\xi_{\mu_1}t^{\nu_2}_{\mu_2}, 
\end{equation}
where $\delta^{\mu_1\cdots\mu_q}_{\nu_1\dots\nu_q}:=q!\delta^{\mu_1}_{[\nu_1}\cdots\delta^{\mu_q}_{\nu_q]}$ is the generalized Kronecker delta and $t^{\mu}_\nu\equiv t_{\rho\nu}g^{\rho\mu}$.
We introduce an auxiliary tensor that has the same algebraic symmetries as the Riemann tensor:
\begin{equation}
    N^{\mu\nu}_{\rho\sigma}=\tfrac{1}{8}\bigl(\delta^{\mu\nu}_{\rho\sigma}+4\lambda^{\mathrm{GB}}\,R^{\mu\nu}_{\rho\sigma}\bigr).
\end{equation}
Then, eq. \eqref{eq:charaEq} can be written in a compact form 
\begin{equation}
     (\mathcal{P}(\xi)\cdot t )^{\mu\nu}=g^{\nu\rho}\delta^{\mu\mu_1\mu_2\mu_3\mu_4}_{\rho\nu_1\nu_2\nu_3\nu_4}N^{\nu_3\nu_4}_{\mu_3\mu_4}\xi^{\nu_1}\xi_{\mu_1}t^{\nu_2}_{\mu_2}. \label{eq:charEqN}
\end{equation}
It follows from the cartoon relations \eqref{eq:ctrelation} that, upon reduction, the independent non-vanishing components of the $N$-tensor are
\begin{equation}
    N^{ab}_{cd}  ,\quad  N^{a w_p}_{bw_q}=N^{aw}_{bw}\delta^{w_p}_{w_q},\quad N^{w_pw_r}_{w_qw_s}=N^{w_1w_2}_{w_1w_2}\delta^{w_pw_r}_{w_qw_s},
\end{equation}
where $a,b\in\{t,x,z\}$, and $p,q,r,s\in\{1,2\}$. In what follows, the early Latin indices $a,b,\ldots,e,f$ run over the directions $\{t,x,z\}$.

Now we consider the cartoon reduction of eq.\eqref{eq:charEqN}. Inserting the non-vanishing polarisations $t_{\mu\nu}\sim(t_{ab},t_{ww})$ and the covector $\xi_\mu\sim(\xi_a,0,0)$,  eq.\eqref{eq:charEqN}  decomposes into a block form: 
\begin{subequations}
     \begin{gather}
     \mathcal{P}^{ac}_{bd}t^{d}_c+ \mathcal{P}^{aw_p}_{bw_p}t^{w}_w=0\,, \label{eq:eqchar1} \\ 
    \mathcal{P}^{wc}_{wd}t^{d}_c+ \mathcal{P}^{ww_p}_{ww_p}t^{w}_w=0\,, 
    \label{eq:eqchar2}
\end{gather} 
\end{subequations}   
where, using the symmetries of the $\mathcal{P}(\xi)$ tensor, we have\footnote{To avoid confusion, we emphasize that the Einstein summation convention is used exclusively for $w_p, w_q, \ldots$, and that $w$ itself is not taken as a dummy index.} 
\begin{gather}
\mathcal{P}^{ac}_{bd}t^{d}_c=\delta^{acew_rw_s}_{bdfw_pw_q}N^{w_pw_q}_{w_rw_s}\xi^{f}\xi_{e}t^{d}_{c}=4N^{w_1w_2}_{w_1w_2}\delta^{ace}_{bdf}\xi^f\xi_et^{d}_c \nonumber\\
    \mathcal{P}^{aw_p}_{bw_p}t^{w}_w=4\delta^{aw_pecw_q}_{bw_pfdw_q}N^{dw}_{cw}\xi^{f}\xi_{e}t^{w}_{w}=8\delta^{ace}_{bdf}N^{dw}_{cw}\xi^f\xi_e t^{w}_w \nonumber\\
     \mathcal{P}^{wc}_{wd}t^{d}_c=4\delta^{wceaw_p}_{wdfbw_p} N^{bw}_{aw}\xi^f\xi_et^d_c=4\delta^{ace}_{bdf}N^{bw}_{aw}\xi^f\xi_e t^{d}_c\nonumber\\  \mathcal{P}^{ww_p}_{ww_p}t^{w}_w=\delta^{eda}_{fcb}N^{cb}_{da}\xi^f\xi_et^w_w.
\end{gather}
Note that \eqref {eq:eqchar1} can be written in a compact form 
\begin{equation}
     (\delta^{ace}_{bdf}\xi^f\xi_e) ~ \bar{t}^{d}_c=0, \label{eq:charGR}
\end{equation}
where $ \bar{t}^{d}_c$ is an auxiliary variable defined as (assuming $N^{w_1w_2}_{w_1w_2}\ne0$\footnote{This assumption holds at weak coupling. More generally, in our coordinates we have $N^{w_1w_2}_{w_1w_2}=\tfrac18\bigl[1+4\lambda^\text{GB}(1-\nabla_aR\,\nabla^aR)/R^2\bigr]$. On the horizon of the UBSs of Section~\ref{sec:UBS_EGB} this equals $\tfrac18(1+4\lambda^\text{GB}/r_H^2)$, which is positive along the entire branch \eqref{eq:range_UBS}. For $\lambda^\text{GB}>0$ it is positive wherever $\nabla_aR\,\nabla^aR\le1$.})
\begin{equation}
    \bar{t}^{d}_c:=t^d_c+\tfrac{2}{N^{w_1w_2}_{w_1w_2}} N^{dw}_{cw} t^w_w.
\end{equation}

A key observation is that eq. \eqref{eq:charGR} is nothing but the principal part of the Einstein equation in $2+1$ dimensions.  We shall now show that $\bar{t}_{ab}$ must be of the form $\xi_{(a}X_{b)}$ for some vector $X_a$. To see this, consider the double Hodge dual of both sides, which gives 
\begin{equation}
    \begin{aligned}        0&=\epsilon_{mka}\epsilon^{b}_{nl}\delta^{ace}_{bdf}\xi^f\xi_e \bar{t}^{d}_c=
    -\xi_f\xi_e \bar{t}_{dc} \delta^{ce}_{mk} \delta^{df}_{nl}\\
   & =
   -\xi_l\xi_k \bar{t}_{nm}+ \xi_n\xi_k \bar{t}_{lm}+
   \xi_l\xi_m \bar{t}_{nk}- \xi_n\xi_m \bar{t}_{lk}, \label{eq:Hudge}
    \end{aligned}
\end{equation}
where, in the second equality, we have repeatedly used the identity $ \delta^{ace}_{bdf}=-\epsilon^{ace}\epsilon_{bdf}$.\footnote{It is worth  noting that the right-hand side of the last equality in \eqref{eq:Hudge} represents the linearized Riemann tensor built from the metric $\bar{t}_{ab}$. This is indeed consistent with the fact that, in $2+1$ dimensions, the Riemann tensor is algebraically fixed by the Einstein tensor.}
By choosing $l=k=x$ and $l=k=z$ in \eqref{eq:Hudge}, we obtain
\begin{subequations}
 \begin{align}
 \xi_x^2 \bar{t}_{mn}&= \xi_{m}V_n+\xi_nV_m ,\label{eq:eqpuregauge1} \\
 \xi_z^2 \bar{t}_{mn}&= \xi_{m}W_n+\xi_n W_m ,\label{eq:eqpuregauge2}
\end{align}
\end{subequations}
where 
\begin{equation}
    \begin{gathered}
        V_n=\bigl(\xi_x\bar{t}_{xn}-\tfrac{1}{2}\xi_n \bar{t}_{xx}\bigr), \\
         W_n=\bigl(\xi_z\bar{t}_{zn}-\tfrac{1}{2}\xi_n \bar{t}_{zz}\bigr).
    \end{gathered}
\end{equation}
Note that, by definition, the spatial part of $\xi$ is non-vanishing, i.e., $\xi_x^2+\xi_z^2\neq0$. Adding \eqref{eq:eqpuregauge1} and \eqref{eq:eqpuregauge2} then implies that any $\bar{t}_{ab}$ satisfying \eqref{eq:charGR} can be written as $\bar{t}_{ab}=\xi_{(a}X_{b)}$.   Consequently, any $t_{ab}$ that solves 
\eqref{eq:eqchar1} must be of the form 
\begin{equation}
    t_{ab}= -\frac{2}{N_{w_1 w_2 w_1w_2}} N_{awbw} t_{ww} +\xi_{(a}X_{b)}
    \label{eq:solutab}
\end{equation}
for some vector $X_a$. This means that $t_{ab}$ is completely determined by $t_{ww}$, up to a high-frequency gauge transformation and, therefore the only physical degree of freedom in the symmetry-reduced system is $t_{ww}$.\footnote{This is intuitive because in this symmetry-reduced setting, at level of the equations of motion the areal radius of the spheres in \eqref{eq:sym_reduced_metric} behaves like a scalar field and in the cartoon reduction this variable is captured by $t_{ww}$}

Substituting \eqref{eq:solutab} back to Eq.\eqref{eq:eqchar2} now gives the the  characteristic equation for $t_{ww}$:
\begin{equation}
\begin{aligned}
     0&= ( 8\delta^{ace}_{bdf}N^{dw}_{cw}N^{bw}_{aw}\xi^f\xi_e -\delta^{eda}_{fcb}N^{w_1w_2}_{w_1w_2}N^{cb}_{da}\xi^f\xi_e) t^w_w\\
     &=\left(8\delta^{ace}_{kdf}N^{dw}_{cw}N^{fw}_{ew}g^{kb}-\delta^{eda}_{fck}N^{w_1w_2}_{w_1w_2}N^{cf}_{de}g^{kb}\right)\xi_a\xi_b t^w_w. 
\end{aligned}
\end{equation}
From this we can read off the effective metric as
\begin{equation}
     g_{\textbf{eff}}^{ab}=\frac{16}{3}g^{kb}\delta^{ace}_{kdf}(8N^{dw}_{cw}N^{fw}_{ew}-N^{df}_{ce}N^{w_1w_2}_{w_1w_2}),
     \label{eq:eff_metric}
\end{equation}
where the constant factor $\tfrac{16}{3}$ is chosen such that $g_{\textbf{eff}}^{ab}$ reduces to the spacetime metric $g^{ab}$ as $\lambda^{\mathrm{GB}}\to0$.

In summary, the non-trivial solution to the characteristic equation \eqref{eq:charaEq} in the symmetry-reduced system is given by the equivalence class  $[t_{\rho\sigma}]$ where
\begin{equation}
    t_{\rho\sigma}=\begin{pmatrix} -\tfrac{2N_{a w b w}}{N_{w_1 w_2 w_1w_2}}\\{1} \end{pmatrix}
    ,  \quad g_{\textbf{eff}}^{ab} ~\xi_a \xi_b=0.
\end{equation}
The effective metric $g_{\textbf{eff}}^{ab}$ derived here controls the propagation of the only physical degree of freedom of the theory when the dynamics is restricted to spherical symmetry in the asymptotically flat directions.

\subsection{ Diagnostic of hyperbolicity loss} 
\label{subsec:hypberb}

As discussed in the preceding subsection \ref{subsec:formulation},  strong hyperbolicity demands that, for any spatial wave vector $\xi_i$,  the root $\xi_0$ of the null cone equation 
\begin{equation}
g_{\textbf{eff}}^{ab}\,\xi_a \xi_b = 0\,,
\end{equation}
must be real. Expanding this equation in powers of $\xi_0$ gives 
\begin{equation}
     g_{\textbf{eff}}^{00} ~ \xi_0^2 +(2g_{\textbf{eff}}^{0i}~\xi_i)\xi_0 +g_{\textbf{eff}}^{ij}~\xi_i\xi_j=0,
\end{equation}
where $i,j\in\{x,z\}$. Then (assuming $g_{\textbf{eff}}^{00}\ne0$), the root $\xi_0$ can be expressed as 
\begin{equation}
    \xi_0^{\pm}=-\frac{g_{\textbf{eff}}^{0i}~\xi_i}{g_{\textbf{eff}}^{00}}\pm \sqrt{\Delta^{ij}\xi_i\xi_j}, \label{eq:disper}
\end{equation}
where we define the \textit{spatially reduced effective metric} $\Delta^{ij}$ as
\begin{equation}
    \Delta^{ij}\equiv -\frac{g_{\textbf{eff}}^{ij}}{g_{\textbf{eff}}^{00}}+\frac{g_{\textbf{eff}}^{0i} ~g_{\textbf{eff}}^{0j}}{(g_{\textbf{eff}}^{00})^2}.
    \label{eq:spatial_eff_metric}
\end{equation}
Note that $\Delta^{ij}$ reduces to $\Delta_{\mathbf{gr}}^{ij}\equiv\alpha^2\gamma^{ij}$ as $\lambda^{\mathrm{GB}}\to0$.
Then, in order for the roots $\xi_0^{\pm}$ to be real, the discriminant $\Delta^{ij}\xi_i\xi_j$ must be non-negative for all $\xi_i$. 

To characterize the local dynamics near the loss of hyperbolicity,  we 
transcribe the dispersion relation \eqref{eq:disper}  into a model equation for a scalar field $u(\tau,x^i)$ in local coordinate space:
\begin{equation}
\partial_\tau^2u=\Delta^{ij}\partial_i\partial_ju, \label{eq:waveEq}
\end{equation}
where $\tau$ is an adapted local time coordinate defined by
\begin{equation}
    \frac{\partial}{\partial\tau}
    \equiv
    \partial_t+\frac{g_{\textbf{eff}}^{0i}}{g_{\textbf{eff}}^{00}}\partial_i.
\end{equation}
Then the dynamical loss of hyperbolicity can be modelled by the change in the character of the model PDE \eqref{eq:waveEq}. 

In $(1+1)$-dimensional systems, the classification of such transitions has been extensively studied \cite{Rassias_1990,Otway:2015aa,Ripley:2019irj,Bernard:2019fjb,R:2022hlf,Doneva:2023oww,Thaalba:2024crk}. A generalization to $2+1$ dimensions is straightforward. 
We begin by diagonalizing the metric $\Delta^{ij}$ via a principal axis transformation, and we denote its largest and smallest eigenvalues by $\Delta_{max}$ and $\Delta_{min}$, respectively.
Suppose the change of character occurs at $\tau = \tau_s$.  
Adopting the terminology used in the $(1+1)$-dimensional case, we classify the transitions into two main types:  
\begin{itemize}
\item[(i)] \textit{Tricomi-type}: if, as $\tau \to \tau_s^{-}$, at least one eigenvalue shrinks to zero. In particular, the  smallest eigenvalue should behave as $ \Delta_{min} \sim \tau_s-\tau$;  
\item[(ii)] \textit{Keldysh-type}: if, as $\tau \to \tau_s^{-}$, at least one eigenvalue diverges.  In particular, the largest eigenvalue should behave as $ \Delta_{max} \sim (\tau_s-\tau)^{-1}$.
\end{itemize}
In both situations, the characteristic speed in some direction becomes imaginary for $\tau>\tau_s$. However, as noted in \cite{Figueras:2020dzx,Bernard:2019fjb}, one should not expect to really see a dynamical change of character in the Keldysh-type transition, because as the characteristic speed blows up, the system becomes extremely stiff, and tends to break down in very short time scale. 

Although we have adopted the Tricomi/Keldysh dichotomy from the $(1+1)$-dimensional case, it should be noted that the $(2+1)$-dimensional system that concerns us admits five possibilities in terms of the asymptotic behaviours of the two eigenvalues at the transition point: (i) $\Delta_{\max}\to\infty$, $0<\Delta_{\min}<\infty$; (ii) $\Delta_{\max},\Delta_{\min}\to\infty$; (iii) $\Delta_{\min}\to 0$, $0<\Delta_{\max}<\infty$; (iv) $\Delta_{\min},\Delta_{\max}\to 0$; and (v) the hybrid case $\Delta_{\min}\to 0$, $\Delta_{\max}\to\infty$. The first two refine the Keldysh-type class, (iii)-(iv) refine the Tricomi-type class, and the final one belongs to the intersection of the two classes. We do not adopt this finer classification here, since the analysis of the data in \cite{Figueras:2026epx} does not allow us to confidently distinguish between these subtypes and this is not the focus of this work.

One might directly consider using the eigenvalues of $\Delta^{ij}$ to diagnose the hyperbolicity loss in the numerical evolutions and  identify the type of  transitions based on the difference in the asymptotic behaviours near the breakdown of the evolution. However, the raw values of $\Delta_{max},\Delta_{min}$ can be extremely large or small even in the GR limit, and the eigenvalues themselves depend on the choice of spatial gauge. Therefore, it is more appropriate to work with some normalized and spatially gauge invariant objects.   To this end, instead of solving the eigenvalue problem for $\Delta^{ij}$, we propose a new pair of diagnostics, namely $\Delta_+$ and $\Delta_-$, defined via the generalized eigenvalue of $\Delta^{ij}$:
\begin{equation}
     \Delta^{ij}\xi_j=\Delta_{\pm}\Delta^{ij}_{\mathbf{gr}}\xi_j \implies \det(\Delta^{ij}-\Delta_{\pm}\Delta^{ij}_{\mathbf{gr}})=0,
\end{equation}
where, as introduced previously, the reference metric $\Delta_{\mathbf{gr}}^{ij}$ corresponds to the value of $\Delta^{ij}$ in the GR limit. Here, $\Delta_{+}$ and  $\Delta_{-}$  are defined to be the largest and the smallest generalized eigenvalues, respectively.
Equivalently, $\Delta_\pm$ are the eigenvalues of the mixed tensor $\frac{1}{\alpha^2}\Delta_{j}^{i}$. By construction, $\Delta_\pm$ are invariant under spatial coordinate changes.

It can be shown that $\Delta_{+}\propto \Delta_{max}$ near the Keldysh-type transition and $\Delta_{-}\propto \Delta_{min}$ near the Tricomi-type transition. Thus, the generalized eigenvalues retain the asymptotic behaviour needed to distinguish between the two types of transition, while providing spatially gauge-invariant diagnostics. In addition, $\Delta_{\pm}$ varies continuously from $1$ in the GR limit to its asymptotic value. Thus, the behaviour of $\Delta_{\pm}$ provides a convenient and quantitative measure of how much the EGB system deviates from GR, and hence of the strength of the coupling.

We can now evaluate the diagnostics $\Delta_\pm$ on the UBS backgrounds of Section~\ref{sec:UBS_EGB}. In static coordinates \eqref{eq:Schw_coords}, the effective metric \eqref{eq:eff_metric} is diagonal and can be written as (no Einstein summation in $a$)
\begin{equation}
    g_{\textbf{eff}}^{ab}=F_a\,g^{ab}\,,\qquad a\in\{t,x,r\}\ \quad(\text{no sum}),
    \label{eq:geff_static}
\end{equation}
where the functions $F_a(r)$ are built out of the curvature of the background and $F_a=1$ for $\lambda^\text{GB}=0$. The $F_a$ are the eigenvalues of the mixed tensor $g_{\textbf{eff}}^{ac}\,g_{cb}$, labelled by the eigenvectors $\partial_t$, $\partial_x$ (the two Killing fields) and the direction orthogonal to them. Therefore, the eigenvalues $F_a$ are scalars and they are continuous across the horizon. In the static slicing the shift vanishes and \eqref{eq:spatial_eff_metric} gives $\Delta^{ij}=\alpha^2\gamma^{ij}F_i/F_t$, so the generalised eigenvalues are simply $F_r/F_t$ and $F_x/F_t$. Using the near-horizon expansion of the UBS solutions, we find that on the horizon
\begin{equation}
\begin{aligned}
    F_t=F_r=&~\sqrt{1-8\,\lambda^\text{GB}/r_H^2}\,,\\
    F_x=&~\sqrt{1-8\,\lambda^\text{GB}/r_H^2}-4\,\lambda^\text{GB}/r_H^2\,,
\end{aligned}
    \label{eq:F_horizon}
\end{equation}
where $r_H$ is the radius of the horizon $S^2$. The equality $F_t=F_r$ on the horizon reflects the fact that the Killing horizon is a characteristic hypersurface of both $g_{\textbf{eff}}^{ab}$ and $g^{ab}$ \cite{Reall:2014pwa,Reall:2021voz,Hollands:2026aeh}. Therefore, on the horizon
\begin{equation}
    \Delta_\pm = \bigg\{1,\,1-\tfrac{4\,\lambda^\text{GB}/r_H^2}{\sqrt{1-8\,\lambda^\text{GB}/r_H^2}}\bigg\}\,.
    \label{eq:Delta_UBS}
\end{equation}
The non-trivial eigenvalue in \eqref{eq:Delta_UBS} is a monotonically decreasing function of $\lambda^\text{GB}/r_H^2$ and it is equal to $1$ in GR. 

For $\lambda^\text{GB}>0$ the non-trivial eigenvalue corresponds to $\Delta_-$, which vanishes when
\begin{equation}
    \tfrac{\lambda^\text{GB}}{r_H^2} = \tfrac{1}{4}(\sqrt{2} -1)\approx 0.104\,, \label{eq:hyperbolicity_bound}
\end{equation}
and becomes negative for larger couplings. This implies that the effective metric changes signature from Lorentzian to Euclidean. The fact that \eqref{eq:Delta_UBS} is evaluated on the horizon does not limit the scope of this result due to the following continuity argument. For $\tfrac14(\sqrt2-1)<\lambda^\text{GB}/r_H^2<\tfrac18$, \eqref{eq:F_horizon} implies that $F_x<0$ on the horizon while $F_t=F_r>0$. Since the $F_a$ are continuous, the same signs persist in an open neighbourhood of the horizon, including a region of the black string exterior. In this region, $\Delta_-=F_x/F_t<0$, so the characteristic speeds \eqref{eq:disper} of the physical mode are imaginary. Since the $F_a$ are scalars, this is a property of the background rather than of the static slicing.

For $\lambda^\text{GB}<0$ the non-trivial generalised eigenvalue corresponds to $\Delta_+$ and satisfies $1<\Delta_+<1+1/\sqrt3$ for any UBS background. Remarkably, this means that the effective metric on the horizon of the UBS never changes signature for negative couplings, even though its curvature invariants diverge as $\lambda^\text{GB}/r_H^2\to-\tfrac14$. 

We conclude that for UBSs with $\tfrac14(\sqrt2-1)<\lambda^\text{GB}/r_H^2<\tfrac18$, the effective metric degenerates on a hypersurface in the exterior (corresponding to $\Delta_-=0$) and becomes elliptic in a region of the exterior ($\Delta_-<0$). Since $\Delta_\pm$ depend continuously on the metric and its first two derivatives, a solution obtained by a well-posed Cauchy evolution from weakly coupled initial data cannot reach the UBSs with $\tfrac14(\sqrt2-1)<\lambda^\text{GB}/r_H^2<\tfrac18$ or converge to them in $C^2$ on compact subsets of the exterior. The same applies to the solutions of the linearized equations on these backgrounds: since the linearized equations share the principal symbol of the non-linear equations, the linearized initial value problem is ill-posed as well. Therefore, the zero modes of Section~\ref{sec:Linear_analysis} with $\lambda^\text{GB}/r_H^2>\tfrac14(\sqrt2-1)$ and any mode solutions of the time-dependent linearized equations on such backgrounds can only be considered formal solutions. In particular, the usual interpretation of the zero modes as the onset of a dynamical instability due to generic perturbations and the predictive meaning of the growth rates are lost in this regime. In the next section we will further investigate this problem by studying the non-linear evolution of the GL instability and the breakdown of Cauchy evolution, and return to the interpretation of these results in Section~\ref{sec:discussion}.

\section{Non-linear evolution and the breakdown of hyperbolicity}
\label{sec:Nonlinear_evo}

In this section we present results of non-linear simulations of the GL instability of black strings in EGB gravity. Here we focus on the characterisation of the breakdown of hyperbolicity; further details of the non-linear dynamics of black strings are discussed in our companion paper \cite{Figueras:2026epx}.

\subsection{Setup and numerical methods}
\label{subsec:NR_setup}

We performed our numerical simulations in the same symmetry reduced setting as in Section~\ref{sec:mccz4_cartoon}, i.e., we assume $SO(3)$ symmetry along the asymptotically flat directions. The symmetry reduction is implemented at the level of the equations of motion with the modified cartoon method \cite{Pretorius:2004jg,Yoshino:2009xp,Cook:2016soy}: we evolve the fields on the half-plane $w_1=w_2=0$, $z\geq0$, with coordinates $(x,z)$, where $x\sim x+L$ is the coordinate along the string and $z$ is the radial coordinate in the asymptotically flat directions. The symmetry-reduced mCCZ4 evolution equations that we solve are given in Appendix~\ref{app:evolution_equations}. Our simulations use the 2D-adapted version of the \texttt{GRChombo} code \cite{Andrade2021}, which was previously used in \cite{Figueras:2022zkg} to study the GL instability of black strings in GR. Moreover, we integrate the main functionality of the \texttt{GRFolres} code \cite{AresteSalo:2023hcp}, which enables us to solve the linear system of the form \eqref{eq:app_cartoon_AK} for $(\partial_t\tilde{A}_{ij},\partial_tK)$.
For further applications of \texttt{GRFolres} to numerical simulations of four-dimensional Einstein-scalar-Gauss-Bonnet gravity see e.g., \cite{AresteSalo:2023mmd,Doneva:2023oww,Doneva:2024ntw,AresteSalo:2025sxc,Corman:2025wun}. 

To allow for a direct comparison with the GR results of \cite{Figueras:2022zkg}, we use the same initial data, constraint damping parameters and mesh refinement scheme as in \cite{Figueras:2022zkg}. In particular, our initial data is a perturbed UBS of GR with thickness $r_0$ in a Kaluza-Klein circle of length $L=10\,r_0$. These initial data solve the constraint equations of GR but not those of EGB gravity. To minimise the initial constraint violations, we start the evolution with $\lambda^\text{GB}=0$ and switch on the coupling as a quadratic function of time until it reaches its target value, which is then kept fixed for the rest of the simulation. Furthermore, to avoid the singular behaviour of the higher derivative terms near the centre of the black string, we follow \cite{AresteSalo:2023mmd} and smoothly switch off the Gauss-Bonnet coupling well inside the AH by replacing
\begin{equation}
    \lambda^\text{GB}\to\frac{\lambda^\text{GB}}{1+e^{-100(\chi-\chi_0)}}\,,\qquad \chi_0=0.1\,,
\end{equation}
where $\chi$ is the conformal factor of the mCCZ4 formulation. The $\chi=0.1$ level set lies well inside the AH \cite{Andrade:2020dgc,Figueras:2022zkg} so this modification should not affect the exterior dynamics.

We have performed simulations with couplings $\lambda^\text{GB}/r_0^2\in\{\pm10^{-4},\pm10^{-5}\}$, as well as GR simulations ($\lambda^\text{GB}=0$) with the mCCZ4 formulation as benchmarks. Since the local coupling is measured by $\lambda^\text{GB}/r_H^2$, where $r_H$ is the local areal radius of the AH along the string, these initial configurations are very weakly coupled. The local relative strength of the GB coupling increases as the string-like segments become thinner during the evolution. The gauge functions in \eqref{eq:eoms_lovelock}--\eqref{eq:eoms_gauge2} are chosen as $a(x)=0$, $b(x)=0.1$ for $\lambda^\text{GB}/r_0^2\in\{0,\pm10^{-5}\}$, and $a(x)=0.001$, $b(x)=0.4$ for $|\lambda^\text{GB}|/r_0^2=10^{-4}$, since the latter runs require a larger separation between the propagation speeds of the unphysical modes and those of the physical ones. While $a(x)$ is generally taken to be non-zero \cite{AresteSalo:2023mmd}, in the symmetry-reduced setting considered here the choice $a(x)=0$ also leads to a strongly hyperbolic evolution system.

To monitor the hyperbolicity of the evolution system, we periodically evaluate the spatially reduced effective metric \eqref{eq:spatial_eff_metric} and its generalised eigenvalues $\Delta_\pm$ on the numerical solution. We compute these quantities everywhere on the computational domain, including inside the AH.

\subsection{Breakdown of hyperbolicity}
\label{subsec:NR_hyperbolicity}

\begin{figure}[t!]
    \centering
    \includegraphics[width=0.9\linewidth]{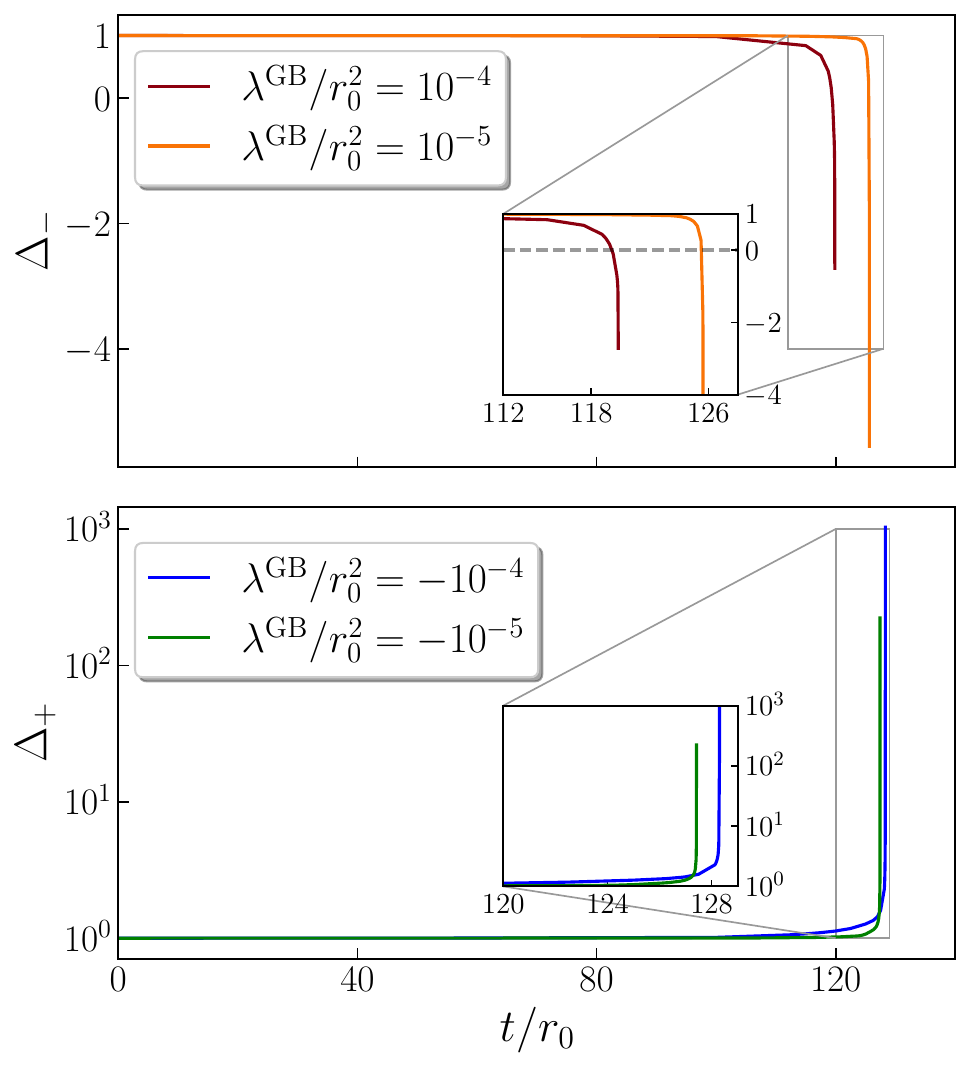}
    \caption{Time evolution of the global minimum of $\Delta_{-}$ for $\lambda^{\mathrm{GB}}/r_0^2=\{10^{-4},10^{-5}\}$ (top) and of the global maximum of $\Delta_{+}$ for $\lambda^{\mathrm{GB}}/r_0^2=\{-10^{-4},-10^{-5}\}$ (bottom). The insets zoom in on the final stages of the evolution. For $\lambda^\text{GB}>0$, $\Delta_-$ crosses zero (Tricomi-type transition) while for $\lambda^\text{GB}<0$ $\Delta_+$ grows by two to three orders of magnitude (Keldysh-type transition).}
    \label{fig:hyper_global}
\end{figure}

\begin{figure*}[t!]
    \centering
    \includegraphics[width=0.48\linewidth]{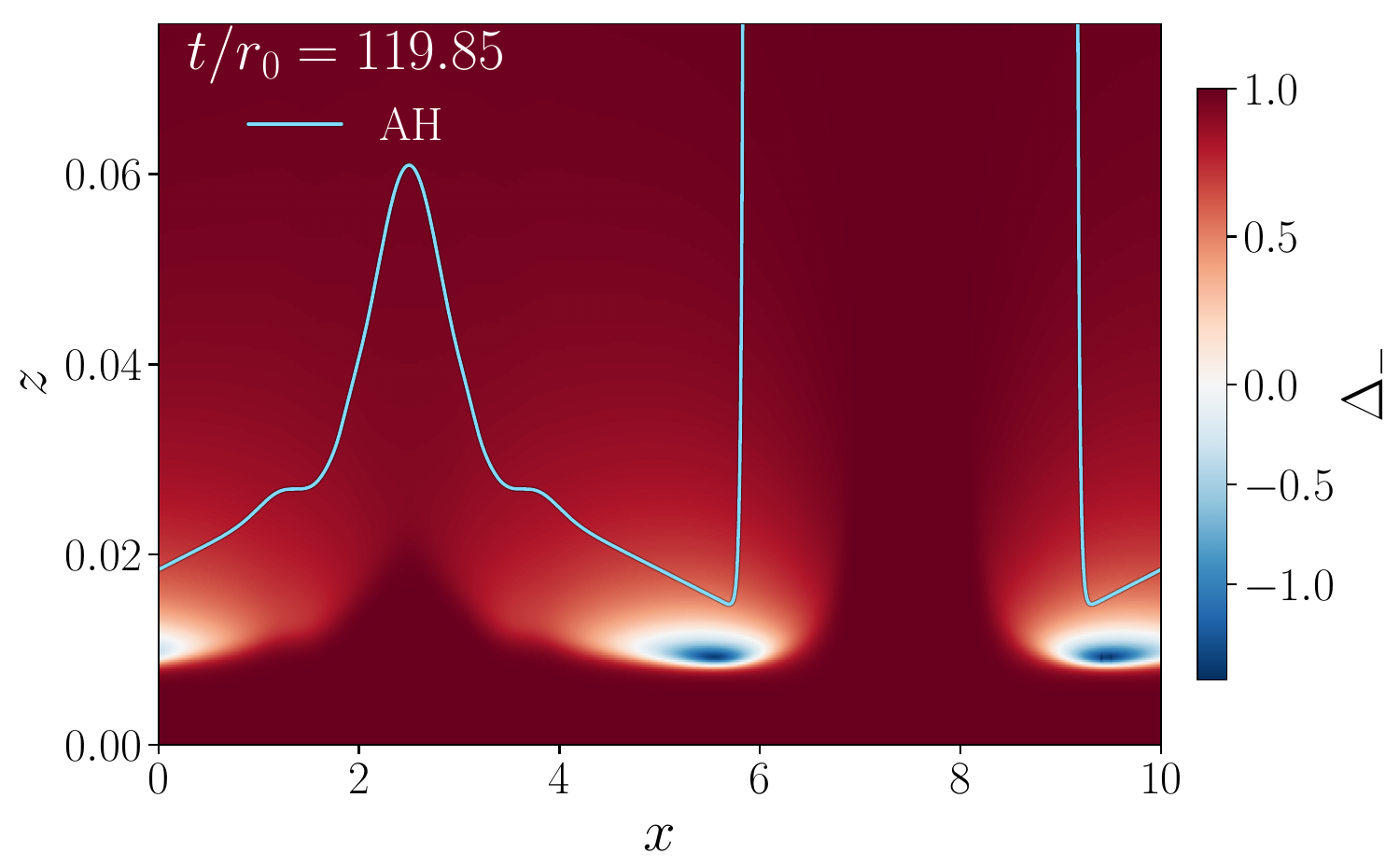}
    \includegraphics[width=0.48\linewidth]{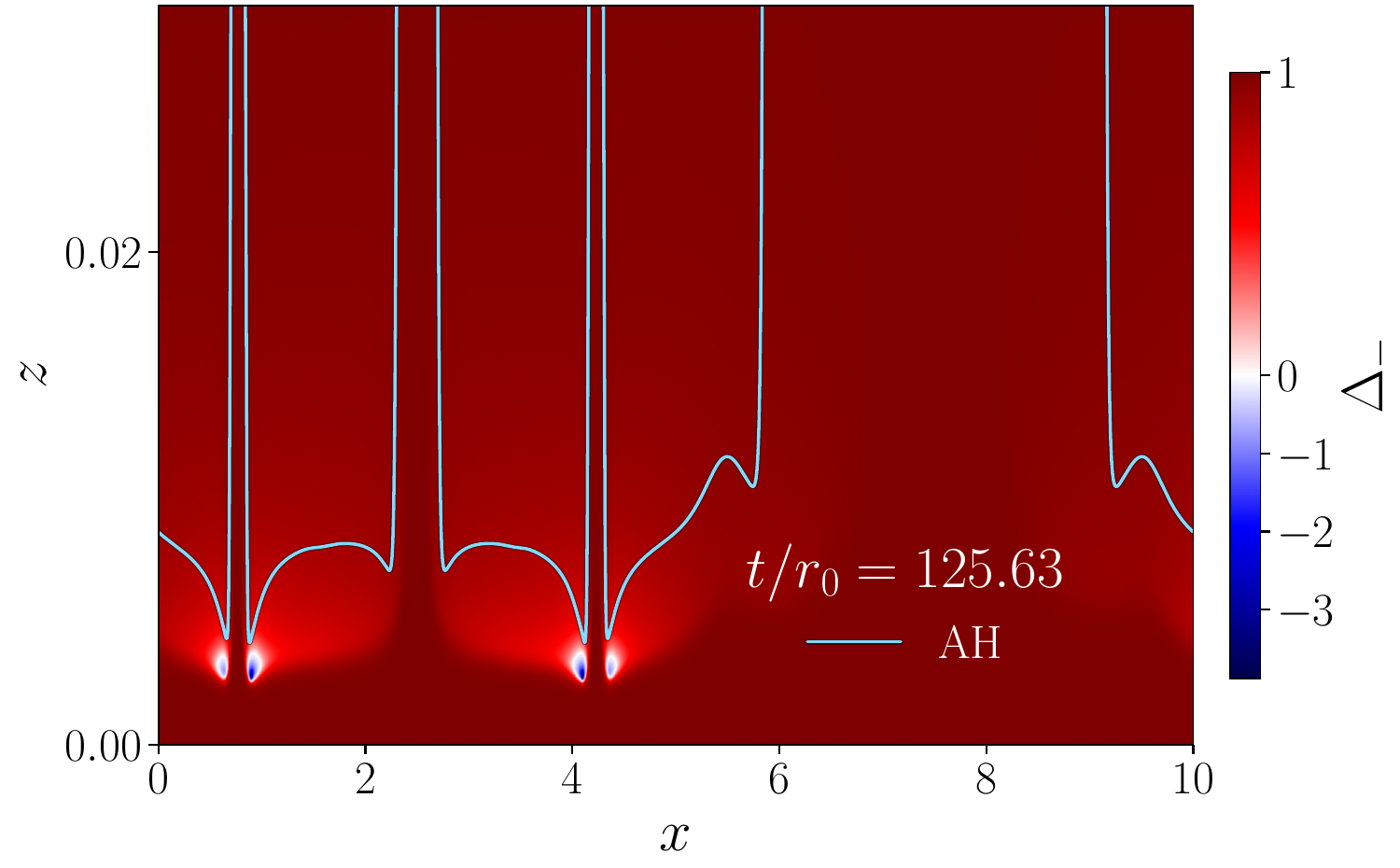}\\
    \includegraphics[width=0.48\linewidth]{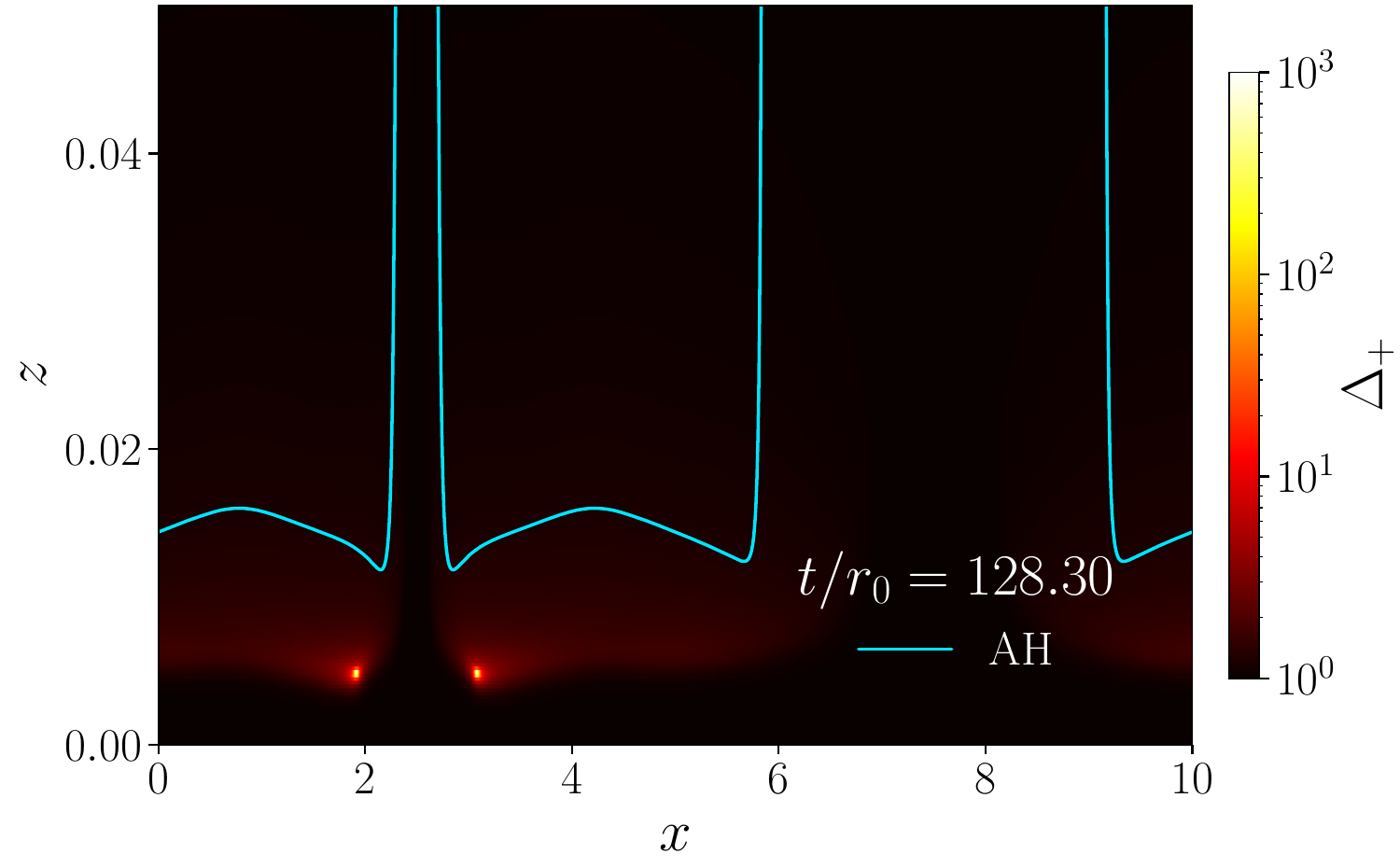}
    \includegraphics[width=0.48\linewidth]{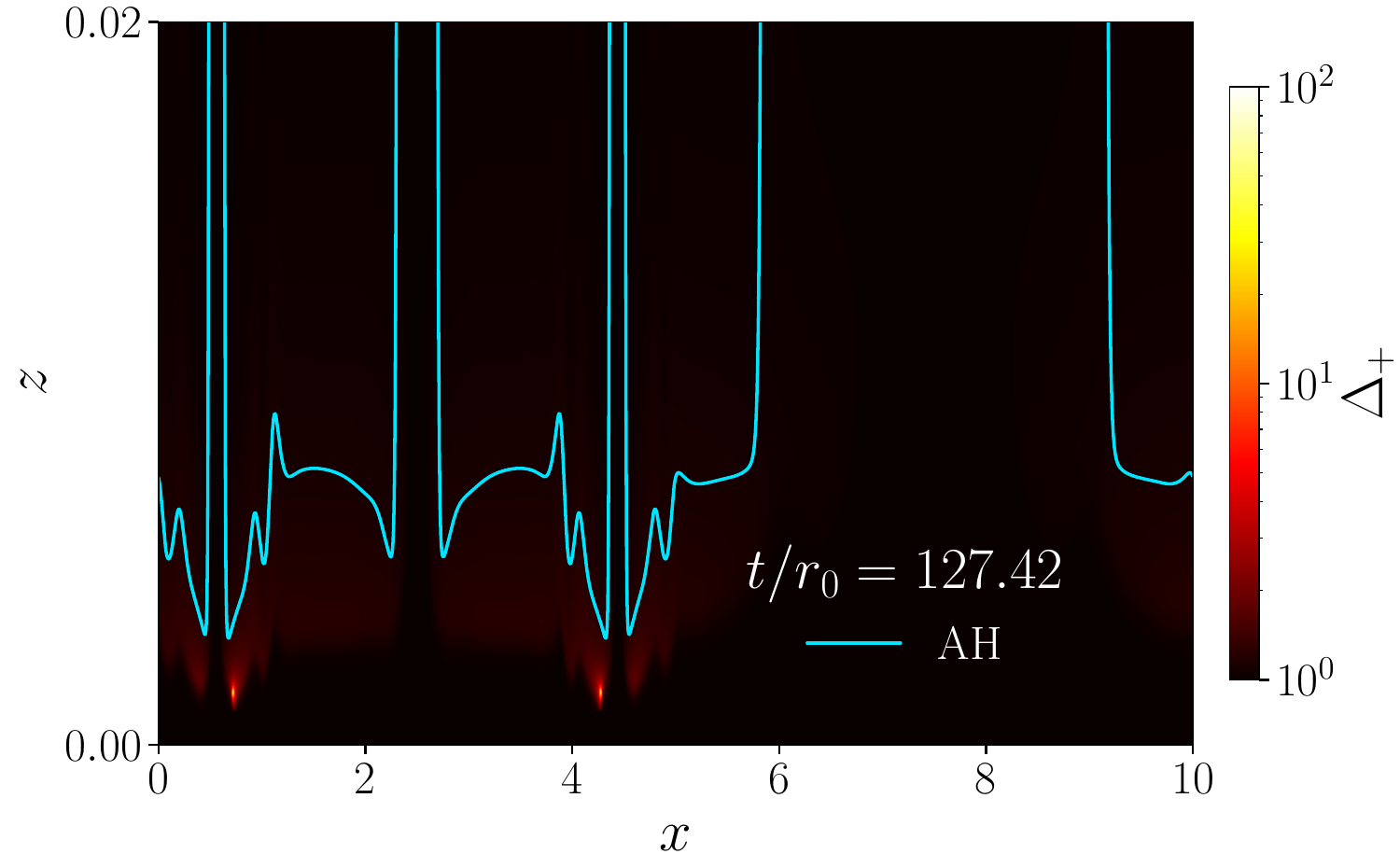}
    \caption{Spatial profiles of $\Delta_\pm$ on the computational domain just before the breakdown of the evolution. \textit{Top}: $\Delta_-$ for $\lambda^\text{GB}/r_0^2=10^{-4}$ (left) and $10^{-5}$ (right). \textit{Bottom}: $\Delta_+$ for $\lambda^\text{GB}/r_0^2=-10^{-4}$ (left) and $-10^{-5}$ (right). The light blue curve is the AH. In all cases the change of character of the equations occurs inside the AH, where the necks join the satellite black holes, i.e., in the thinnest regions of the string.}
    \label{fig:hyper_profiles}
\end{figure*}

For all values of the coupling that we have considered, the system remains weakly coupled for most of the evolution. As a result, $\Delta_\pm$ stay very close to $1$ everywhere, see Fig.~\ref{fig:hyper_global}. Significant deviations from GR only appear at late times (for either sign of the GB coupling) when the thinnest string segments have become very thin, leading to rapid changes in $\Delta_\pm$ and the termination of the simulations due to a loss of hyperbolicity. The nature of the breakdown depends on the sign of $\lambda^\text{GB}$.

For $\lambda^\text{GB}>0$, the global minimum of $\Delta_-$ decreases steeply at late times and crosses zero, signalling a Tricomi-type transition (top panel of Fig.~\ref{fig:hyper_global}). Beyond this point, there is a region of the computational domain in which $\Delta_-<0$, so the characteristic speed of the physical mode is imaginary in some direction and the initial value problem for subsequent evolution becomes ill-posed. The spatial profiles in the top row of Fig.~\ref{fig:hyper_profiles} show that the region where $\Delta_-<0$ first appears inside the AH is in the thinnest part of the string where the necks join the satellite black holes. This is also the region where the curvature invariants on the AH are largest and the string is most non-uniform \cite{Figueras:2026epx}. At this time, $\Delta_-$ remains positive on and outside the AH, although it deviates from $1$ by an amount of order unity near the thinnest segments of the string.

For $\lambda^\text{GB}<0$, the global maximum of $\Delta_+$ grows by two to three orders of magnitude within a short time interval at the end of the evolution (bottom panel of Fig.~\ref{fig:hyper_global}), which indicates that the characteristic speed of the physical mode diverges and a Keldysh-type transition takes place. As discussed in Section~\ref{subsec:hypberb}, one should not expect to see a dynamical change of character in this case since the system becomes extremely stiff \cite{Figueras:2020dzx,Bernard:2019fjb}. As in the $\lambda^\text{GB}>0$ case, the bottom row of Fig.~\ref{fig:hyper_profiles} shows that the transition is localised inside the AH, near the thinnest necks joining the blobs, although the points where $\Delta_+$ is largest lie deeper inside the AH than the region where $\Delta_-<0$ in the $\lambda^\text{GB}>0$ case. Recall that the Gauss-Bonnet term is effectively switched off further inside the AH so the transition cannot occur arbitrarily close to the centre of the string. An analogous dependence of the type of transition on the sign of the coupling has been observed in scalar \cite{Figueras:2025wtx} and scalar-tensor theories \cite{Bernard:2019fjb}.

\subsection{Comparison with the UBS hyperbolicity threshold}
\label{subsec:NR_vs_UBS}

It is natural to compare the breakdown of the non-linear evolution for $\lambda^\text{GB}>0$ with the hyperbolicity threshold \eqref{eq:hyperbolicity_bound} evaluated on the near-horizon region of UBS backgrounds. At the breakdown of the $\lambda^\text{GB}/r_0^2=10^{-5}$ simulation, we have $R_\text{min}\approx0.011\,r_0\approx3.5\sqrt{\lambda^\text{GB}}$ which corresponds to a local coupling $\lambda^\text{GB}/R_\text{min}^2\approx0.08$. This is below the threshold value $\tfrac14(\sqrt2-1)\approx0.104$ obtained from the UBS background. This means that the thinnest segment of the AH at the breakdown is still about $15\%$ thicker than the marginal UBS.

However, we emphasise that this comparison is only heuristic, since the string-like segments are highly non-uniform at late times and are not accurately described by UBSs \cite{Figueras:2026epx}. Furthermore, \eqref{eq:Delta_UBS} concerns the horizon of the UBS, while the loss of hyperbolicity in the simulations occurs inside the AH where the curvature is larger. However, both the earlier onset of the transition and its location are consistent with the picture that emerges from Section~\ref{subsec:hypberb}: the UBSs with $\lambda^\text{GB}/r_H^2>\tfrac14(\sqrt2-1)$, in particular the regular UBS of minimal thickness, cannot be approached in the non-linear evolution of the GL instability.

For $\lambda^\text{GB}<0$, the comparison with UBSs is even more striking. At the breakdown of the $\lambda^\text{GB}/r_0^2=-10^{-5}$ simulation we have $R_\text{min}\approx0.009\,r_0$, corresponding to a local coupling $\lambda^\text{GB}/R_\text{min}^2\approx-0.13$. This is quite far from the limiting value $-\tfrac14$. Note, however, that this is consistent with the fact that in our simulations with $\lambda^\text{GB}<0$, the breakdown of hyperbolicity happens further inside the AH.

\section{Discussion}
\label{sec:discussion}

In GR, the well-posedness of the initial value problem guarantees that, for small enough perturbations, solutions of the linearized equations approximate solutions of the full theory. Therefore, linear (in)stability analyses directly inform the non-linear dynamics. In EGB gravity, however, the equations of motion are only guaranteed to be hyperbolic in a suitably weakly coupled regime, whereas both the family of UBSs and the linear perturbations around them extend into the strongly coupled regime. This means that the implicit assumption that solutions to the linearized equations around highly symmetric backgrounds are meaningful approximations to solutions of the non-linear equations is not always true in Einstein-Gauss-Bonnet theory, and more generally, in effective field theories applied outside of their regime of validity!

It is worth further discussing and contrasting this interpretation of loss of hyperbolicity with the way in which it is often described in studies of linear perturbations of black holes and black branes in Lovelock theories. In several works \cite{Gleiser:2005ra,Beroiz:2007gp,Takahashi:2010gz,Buchel:2009tt,Konoplya:2017ymp,Buchel:2026rep}, the failure of hyperbolicity of the linearized equations in some region of a stationary background is associated with the appearance of short wavelength ``instabilities'' in the spectrum of quasinormal modes whose growth rates increase without bound with the wavenumber or angular momentum number. As emphasised in \cite{Reall:2014pwa,Papallo:2017qvl}, such modes should not be regarded as a physical instability of the background. A genuine linear instability, such as the GL instability studied in Section~\ref{sec:Linear_analysis}, has a growth rate that is bounded from above so that the norm of a generic perturbation grows at most exponentially and at a rate set by the background. Growth rates that are unbounded in the wavenumber are instead manifestations of the ill-posedness of the initial value problem: in any fixed time, high frequency perturbations of the initial data that are arbitrarily small in any Sobolev norm are amplified by an arbitrarily large factor so solutions do not depend continuously on the initial data. Since the linearized equations share the principal symbol of the non-linear ones, the same conclusion applies to the full theory: in any region in which $g_{\textbf{eff}}^{ab}$ fails to be Lorentzian, the Cauchy problem for EGB gravity is ill-posed both within the symmetry class considered here and also in general. This means that for generic initial data one should not expect a solution to exist at all. In particular, we expect that exact stationary solutions such as the UBSs with $\tfrac14(\sqrt2-1)<\lambda^\text{GB}/r_H^2\le\tfrac18$ cannot arise as the Cauchy development of an initial data set. This is consistent with the non-linear evolutions of \cite{Figueras:2026epx}. We therefore do not interpret the loss of hyperbolicity identified via $\Delta_\pm$ as a new instability of black strings in EGB gravity, but as the point beyond which the truncated theory \eqref{eq:action5d} no longer provides a predictive classical description. Whether the evolution can be continued in the sense of EFT (e.g., by using the regularisation method of \cite{Figueras:2024bba}) is a separate question that we leave for future work.

We also emphasise that the bound \eqref{eq:hyperbolicity_bound} has been derived within the $SO(3)$-symmetric sector which contains a single physical degree of freedom. Perturbations that break this symmetry can propagate along a larger set of characteristic hypersurfaces of the full principal symbol, which may fail to be hyperbolic at smaller values of the coupling \cite{Reall:2014pwa}. Therefore \eqref{eq:hyperbolicity_bound} provides an upper bound on the range of couplings for which EGB gravity admits a well-posed initial value problem in a neighbourhood of the UBS horizon. UBSs with $\lambda^\text{GB}/r_H^2>\tfrac14(\sqrt2-1)$ certainly lie outside this range, and the exact threshold requires an analysis of all physical polarisations. However, this may be a challenging task since Eisntein-Gauss-Bonnet theory in 5 dimensions admits 5 gravitational polarisations so characteristic surfaces are determined by a polynomial of degree 10, which is not expected to factorise to quadratic cones on a generic background.

It is interesting to note that in an EGB theory with $\lambda^\text{GB}<0$, the effective metric remains Lorentzian on and outside the horizon of any uniform black string. This, together with the fact that the connecting string segments remain fairly uniform even during the non-linear evolution in the $\lambda^\text{GB}<0$ case, raises the possibility of extending the non-linear simulations, approaching the singular threshold UBS solution more closely and thus producing even larger curvature invariants in the exterior of the black string. Doing so would provide more evidence for the violation of weak cosmic censorship in this setting. This would likely involve a more aggressive excision of the interior of the AH and handling the stiffness of the evolution equations in the near-horizon region.

\section*{Acknowledgements}
PF would also like to thank the Enrico Fermi Institute and Department of Physics of the University of Chicago, where part of this work was carried out, for their hospitality. We also thank the GRTL collaboration (\texttt{www.grtlcollaboration.org}) for their support and code development work. PF and ADK are supported by the STFC Consolidated Grant ST/X000931/1  (Astronomy at Queen Mary 2023-2026). SY was supported by the China Scholarship Council.  The authors acknowledge the use of the Sulis HPC facility and the support of the Sulis Research Software Engineer (RSE) team. Sulis is funded by the Engineering and Physical Sciences Research Council (EPSRC) (grant number EP/T022108/1) and the HPC Midlands+ consortium. This work also used the DiRAC@Durham facility managed by the Institute for Computational Cosmology on behalf of the STFC DiRAC HPC Facility (www.dirac.ac.uk). The equipment was funded by BEIS capital funding via STFC capital grants ST/P002293/1, ST/R002371/1 and ST/S002502/1, Durham University and STFC operations grant ST/R000832/1. DiRAC is part of the National e-Infrastructure. We also acknowledge the EuroHPC Joint Undertaking for awarding this project access to the EuroHPC supercomputer LUMI-C, hosted by CSC (Finland) and the LUMI consortium through a EuroHPC Regular Access call grant number EHPC-REG-2025R01-217. Finally, the Texas Advanced Computing Center (TACC) at The University of Texas at Austin also provided computational resources that have contributed to the research results reported in this paper. URL: http://www.tacc.utexas.edu

\appendix

\section{Spherical gauge fixing}
\label{app:gauge_fixing}
In this appendix we show that a general metric perturbation $h_{\mu\nu}$ of the background UBS in the coordinates \eqref{eq:Schw_coords} can be brought into the spherical gauge \eqref{eq:spherical_gauge}.

Consider the general, time- and $x$-dependent metric perturbations that preserve the spherical symmetry (in the asymptotically flat directions) of the background UBS \eqref{eq:Schw_coords}:
\begin{equation}
    \begin{aligned}
        h_{\mu\nu}\,dx^\mu\,dx^\nu=&~e^{\Omega t}\Bigl\{
            \cos(k\,x)\bigl[ 
                H_{tt}(r)\,dt^2+2\,H_{tr}(r)\,dt\,dr\\
                &\hspace{1.5cm}+H_{rr}(r)\,dr^2+H_S(r)\,d\Omega_{(2)}^2 \\
                &\hspace{1.5cm}+ H_{xx}(r)\,dx^2
            \bigr]\\
            &\hspace{0.2cm}+2\,\sin(k\,x)\bigl[ 
                H_{tx}(r)\,dt\,dx+H_{rx}(r)\,dr\,dx
            \bigr]
        \Bigr\}
    \end{aligned}
    \label{eq:hab_general}
\end{equation}
Similarly, consider the general gauge parameter compatible with the symmetries of the problem:
\begin{equation}
    \begin{aligned}
    \xi_\mu\,dx^\mu=&~e^{\Omega t}\bigl[ 
        \cos(k\,x)\bigl( \xi_t(r)\,dt+\xi_r(r)\,dr\bigr) \\
        &\hspace{0.75cm}+ \sin(k\,x)\,\xi_x(r)\,dx
    \bigr]\,.
    \end{aligned}
\end{equation}

Making a gauge transformation, 
\begin{equation}
    h_{\mu\nu}\to h_{\mu\nu} + 2\,\nabla_{(\mu}\xi_{\nu)}\,,
\end{equation}
with $k\neq 0$ and $\Omega\neq0$, we can choose
\begin{equation}
    \begin{aligned}
        \xi_r =& -\tfrac{B(r)}{S'(r)}\,H_S(r)\,,\\
        \xi_x =&\frac{-H_{tx}(r)+k\,\xi_t}{\Omega}\,,
    \end{aligned}
\end{equation}
and
\begin{equation}
    \begin{aligned}
        \xi_t=&~\tfrac{A(r)\,W(r)}{W(r)A'(r)-A(r)W'(r)}\Bigl[ 
            H_{tr}(r) - \tfrac{2\,\Omega\,B(r)}{S'(r)}\,H_S(r) \\
            &\hspace{0.4cm}-\tfrac{\Omega}{k}\,H_{rx}(r) + \tfrac{1}{k}\Bigl( H'_{tx}(r)-\tfrac{W'(r)}{W(r)}\,H_{tx}(r)\Bigr)
        \Bigr]\,,
    \end{aligned}
\end{equation}
to bring the metric perturbations into the form \eqref{eq:spherical_gauge}.

\section{Cartoon reduction of the evolution equations}
\label{app:evolution_equations}

We evolve Einstein-Gauss-Bonnet gravity using the modified CCZ4 formulation in $d+1$ dimensions of Ref.~\cite{AresteSalo:2023mmd}, to which we refer for the full $d+1$ equations; the ones implemented in our code differ from those only in the normalisation of the Gauss-Bonnet terms.\footnote{In Ref.~\cite{AresteSalo:2023mmd} the Gauss-Bonnet contributions enter through the effective matter terms $\kappa\rho$, $\kappa J_i$, $\kappa\bar S^{\text{TF}}_{ij}$ and $\kappa\bar S$, into which the coupling $\lambda^{\rm GB}$ is absorbed, and the quantities $\hat{\mathcal O}_{ij}$ and $\hat{\mathcal O}$ defined there correspond to our $\hat{\mathcal O}_{ij}/\alpha$ and $\hat{\mathcal O}/\alpha$. In terms of the quantities used here, $\kappa\rho=\lambda^{\rm GB}\rho^{GB}$, $\kappa J_i=\lambda^{\rm GB}J^{GB}_i$, $\kappa\bar S^{\text{TF}}_{ij}=-\lambda^{\rm GB}\tilde S^{GB}_{ij}/(\alpha\chi)$ and $\kappa\bar S=(d-1)\lambda^{\rm GB}S^{GB}_K/\alpha+d\,\lambda^{\rm GB}\rho^{GB}$.} Here we only give their cartoon reduction for the spatial geometry with an $SO(d-1)$ symmetry,
\begin{equation}
ds^2=g_{ab}(x)\,dx^a\,dx^b+R(x)^2\,d\Omega^2_{(d-2)}\,,
\end{equation}
where $d\Omega^2_{(d-2)}$ is the metric on the round unit $(d-2)$-sphere. The indices $a,b,\ldots$ run over the Cartesian coordinates $(x,z)$ of the computational domain, $w$ labels any of the $n\equiv d-2$ sphere directions (with $w_p$, $p=1,\ldots,n$, when they need to be distinguished), and all quantities are evaluated on the $z$ axis of the extra dimensions. We use the conformal variables $\{\chi,\tilde\gamma_{ij},K,\tilde A_{ij},\Theta,\hat\Gamma^i\}$ of Ref.~\cite{AresteSalo:2023mmd} and $\partial_0\equiv\partial_t-\beta^c\partial_c$. The indices of $Z_i$ and $\tilde A_{ij}$ are raised and lowered with $\tilde\gamma_{ij}$, while in the Gauss-Bonnet terms all other indices are raised and lowered with the physical metric $\gamma_{ij}$. The non-vanishing components of the Christoffel symbols of $\tilde\gamma_{ij}$ are
\begin{subequations}
\label{eq:app_cartoon_Gamma}
\begin{align}
\tilde\Gamma^a_{\phantom{a}bc} &= \tfrac{1}{2}\,\tilde\gamma^{ad}\big(\partial_b\tilde\gamma_{cd}+\partial_c\tilde\gamma_{bd}-\partial_d\tilde\gamma_{bc}\big)\,, \label{eq:app_Gamma_abc}\\
\tilde\Gamma^a_{\phantom{a}w_pw_q} &= \delta_{pq}\,\tilde\Gamma^a_{\phantom{a}ww}\,, \label{eq:app_Gamma_awpwq}\\
\tilde\Gamma^a_{\phantom{a}ww} &= \tfrac{1}{z}\big(\delta_z^{\phantom{z}a}-\tilde\gamma^{az}\tilde\gamma_{ww}\big)-\tfrac{1}{2}\,\tilde\gamma^{ab}\partial_b\tilde\gamma_{ww}\,, \label{eq:app_Gamma_ww}\\
\tilde\Gamma^{w_p}_{\phantom{w_p}w_qa} &= \delta^p_q\,\tilde\Gamma^w_{\phantom{w}wa}\,, \label{eq:app_Gamma_wpwqa}\\
\tilde\Gamma^w_{\phantom{w}wa} &= \tfrac{1}{2}\,\tilde\gamma^{ww}\partial_a\tilde\gamma_{ww}\,, \label{eq:app_Gamma_wa}
\end{align}
\end{subequations}
and we define
\begin{equation}
\text{div}\beta\equiv\partial_i\beta^i=\partial_a\beta^a+\tfrac{n}{z}\,\beta^z\,.
\end{equation}

\paragraph{Reduced evolution equations.}
The $d+1$ equations of Ref.~\cite{AresteSalo:2023mmd} (Eqs.~(14) and (C4) there) reduce to
\begin{subequations}
\label{eq:app_cartoon_evol}
\begin{align}
&\partial_0\chi = \tfrac{2}{d}\,\chi\big(\alpha\,K-\text{div}\beta\big)\,,\\
&\partial_0\tilde\gamma_{ab} = -2\,\alpha\,\tilde A_{ab}+2\,\tilde\gamma_{c(a}\partial_{b)}\beta^c-\tfrac{2}{d}\,\tilde\gamma_{ab}\,\text{div}\beta\,,\\
&\partial_0\tilde\gamma_{ww} = -2\,\alpha\,\tilde A_{ww}-\tfrac{2}{d}\,\tilde\gamma_{ww}\big(\text{div}\beta-\tfrac{d}{z}\,\beta^z\big)\,,\\
&\partial_0\Theta = \tfrac{\alpha}{2}\Big[R-\tilde A_{ab}\tilde A^{ab}-n\Big(\frac{\tilde A_{ww}}{\tilde\gamma_{ww}}\Big)^2+\tfrac{d-1}{d}\,K^2 \nonumber\\
&\qquad+2\,D^aZ_a-2\,\Theta\,K\Big]-Z_c\,D^c\alpha-\lambda^{\rm GB}\,\alpha\,\rho^{GB} \nonumber\\
&\qquad-\tfrac{\kappa_1}{2}\big(d+1+(d-1)\kappa_2\big)\,\alpha\,\Theta \nonumber\\
&\qquad-\tfrac{b(x)}{1+b(x)}\Big\{\tfrac{\alpha}{2}\,H-\tfrac{1}{d-1}\,Z_c\,D^c\alpha-\lambda^{\rm GB}\,\alpha\,\rho^{GB} \nonumber\\
&\qquad\quad-\tfrac{\kappa_1}{2}\big[\tfrac{d-3}{2+b(x)}+d+1+(d-1)\kappa_2\big]\,\alpha\,\Theta\Big\}\,,\\
&\partial_0\hat\Gamma^a = -2\,\tilde A^{ab}\partial_b\alpha+2\,\alpha\big(\tilde\Gamma^a_{\phantom{a}bc}\tilde A^{bc}+n\,\tilde\Gamma^a_{\phantom{a}ww}\tilde A^{ww} \nonumber\\
&\qquad-\tfrac{d-1}{d}\,\tilde\gamma^{ab}\partial_bK-\tfrac{d}{2\chi}\,\tilde A^{ab}\partial_b\chi\big)-\hat\Gamma^b\partial_b\beta^a+\tfrac{2}{d}\,\hat\Gamma^a\,\text{div}\beta \nonumber\\
&\qquad+\tfrac{d-2}{d}\,\tilde\gamma^{ac}\Big[\partial_c\partial_b\beta^b+n\Big(\frac{\partial_c\beta^z}{z}-\delta_c^{\phantom{c}z}\frac{\beta^z}{z^2}\Big)\Big] \nonumber\\
&\qquad+\tilde\gamma^{cd}\partial_c\partial_d\beta^a+\frac{n}{\tilde\gamma_{ww}}\Big(\frac{\partial_z\beta^a}{z}-\delta_z^{\phantom{z}a}\frac{\beta^z}{z^2}\Big) \nonumber\\
&\qquad+2\,\alpha\,\tilde\gamma^{ab}\big(\partial_b\Theta-\tfrac{1}{\alpha}\,\Theta\,\partial_b\alpha-\tfrac{2}{d}\,K\,Z_b\big) \nonumber\\
&\qquad-2\,\kappa_1\,\alpha\,\tilde\gamma^{ab}Z_b-2\,\lambda^{\rm GB}\,\alpha\,\tilde\gamma^{ab}J^{GB}_b \nonumber\\
&\qquad-\tfrac{2\,b(x)}{1+b(x)}\,\alpha\Big[\tilde\gamma^{ab}N_b+\tilde\gamma^{ab}\big(\partial_b\Theta-\tfrac{1}{d}\,K\,Z_b\big) \nonumber\\
&\qquad\quad-\tilde A^{ab}Z_b-\kappa_1\,\tilde\gamma^{ab}Z_b-\lambda^{\rm GB}\,\tilde\gamma^{ab}J^{GB}_b\Big]\,,
\end{align}
\end{subequations}
together with
\begin{align}
\arraycolsep=2pt
&\left(
\begin{array}{ccc}
X_{ab}^{\phantom{ab}cd} & n\,X_{ab}^{\phantom{ab}ww} & Y_{ab} \\
X_{ww}^{\phantom{ww}cd} & X_{ww}^{\phantom{ww}ww} & Y_{ww} \\
X^{cd} & n\,X^{ww} & Y
\end{array}
\right)
\left(
\begin{array}{c}
\partial_t\tilde A_{cd} \\
\partial_t\tilde A_{ww}\\
\partial_t K
\end{array}
\right) 
=
\left(
\begin{array}{c}
Z_{ab} \\
Z_{ww}\\
Z
\end{array}
\right)\,,
\label{eq:app_cartoon_AK}
\end{align}
where $\tilde A^{ww}=\tilde\gamma_{ww}^{-2}\,\tilde A_{ww}$, the term $Z_cD^c\alpha\equiv\gamma^{cd}Z_c\partial_d\alpha$ is computed with the physical metric, and $H\equiv M$ and $N_a$ are the Hamiltonian and momentum constraints given below.

\paragraph{Derivatives of the lapse.}
We have
\begin{subequations}
\label{eq:app_boxalpha}
\begin{align}
D^iD_i\alpha&=\chi\Big[\tilde D^i\tilde D_i\alpha-\tfrac{d-2}{2\chi}\,\tilde\gamma^{ab}\,\partial_a\alpha\,\partial_b\chi\Big]\,,\\
\tilde D^i\tilde D_i\alpha&=\tilde\gamma^{ab}\partial_a\partial_b\alpha-\tilde\Gamma^c\partial_c\alpha+\frac{n}{z\,\tilde\gamma_{ww}}\,\partial_z\alpha\,,
\end{align}
\end{subequations}
where $\tilde\Gamma^a$ contains contributions from the sphere directions,
\begin{equation}
\tilde\Gamma^a=\tilde\gamma^{bc}\,\tilde\Gamma^a_{\phantom{a}bc}+\frac{n}{\tilde\gamma_{ww}}\,\tilde\Gamma^a_{\phantom{a}ww}\,,
\end{equation}
and
\begin{subequations}
\begin{align}
&(D_aD_b\alpha)^{\text{TF}} = \tilde D_a\tilde D_b\alpha+\tfrac{1}{\chi}\,\partial_{(a}\alpha\,\partial_{b)}\chi \nonumber\\
&\qquad-\tfrac{1}{d}\,\tilde\gamma_{ab}\big(\tilde D^i\tilde D_i\alpha+\tfrac{1}{\chi}\,\tilde\gamma^{cd}\,\partial_c\alpha\,\partial_d\chi\big)\,,\\
&(D_wD_w\alpha)^{\text{TF}} = \tfrac{1}{z}\,\tilde\gamma_{ww}\,\tilde\gamma^{za}\partial_a\alpha+\tfrac{1}{2}\,\tilde\gamma^{ab}\,\partial_a\tilde\gamma_{ww}\,\partial_b\alpha \nonumber\\
&\qquad-\tfrac{1}{d}\,\tilde\gamma_{ww}\big(\tilde D^i\tilde D_i\alpha+\tfrac{1}{\chi}\,\tilde\gamma^{cd}\,\partial_c\alpha\,\partial_d\chi\big)\,.
\end{align}
\end{subequations}

\paragraph{Ricci tensor.}
We split $R_{ij}=\tilde R_{ij}+R^\chi_{ij}$ and $D_iZ_j=\tilde D_iZ_j+D^\chi_iZ_j$, with
\begin{subequations}
\begin{align}
\tilde R_{ij} =&~ -\tfrac{1}{2}\,\tilde\gamma^{kl}\partial_k\partial_l\tilde\gamma_{ij}+\tilde\gamma_{k(i}\partial_{j)}\tilde\Gamma^k-\partial_k\tilde\gamma_{l(i}\,\partial_{j)}\tilde\gamma^{kl} \nonumber\\
&+\tfrac{1}{2}\,\tilde\Gamma^k\partial_k\tilde\gamma_{ij}-\tilde\Gamma^k_{\phantom{k}li}\,\tilde\Gamma^l_{\phantom{l}kj}\,,\\
R^\chi_{ij} =&~ \tfrac{d-2}{2\chi}\Big[\tilde D_i\tilde D_j\chi-\tfrac{1}{2\chi}\,\partial_i\chi\,\partial_j\chi\Big] \nonumber\\
&+\tfrac{1}{2\chi}\,\tilde\gamma_{ij}\Big[\tilde D_k\tilde D^k\chi-\tfrac{d}{2\chi}\,\tilde\gamma^{kl}\,\partial_k\chi\,\partial_l\chi\Big]\,,\\
D^\chi_iZ_j =&~ \tfrac{1}{\chi}\Big[Z_{(i}\,\partial_{j)}\chi-\tfrac{1}{2}\,\tilde\gamma_{ij}\,\tilde\gamma^{kl}Z_k\,\partial_l\chi\Big]\,.
\end{align}
\end{subequations}
The components entering \eqref{eq:app_cartoon_evol} and \eqref{eq:app_cartoon_AK} are then
\begin{subequations}
\begin{align}
&\tilde R_{ab}+2\,\tilde D_{(a}Z_{b)} = -\tfrac{1}{2}\,\tilde\gamma^{cd}\partial_c\partial_d\tilde\gamma_{ab}+\tilde\gamma_{c(a}\partial_{b)}\hat\Gamma^c \nonumber\\
&\quad-\partial_c\tilde\gamma_{d(a}\,\partial_{b)}\tilde\gamma^{cd}+\tfrac{1}{2}\,\hat\Gamma^c\partial_c\tilde\gamma_{ab}-\tilde\Gamma^c_{\phantom{c}da}\,\tilde\Gamma^d_{\phantom{d}cb} \nonumber\\
&\quad+n\Big\{-\tfrac{1}{2\tilde\gamma_{ww}}\Big[\tfrac{1}{z}\,\partial_z\tilde\gamma_{ab}-\tfrac{2}{z^2}\big(\tilde\gamma_{z(a}\,\delta_{b)z}-\tilde\gamma_{ww}\,\delta_{az}\,\delta_{bz}\big)\Big] \nonumber\\
&\qquad+\tfrac{1}{z\,\tilde\gamma_{ww}^2}\big(\tilde\gamma_{z(a}-\tilde\gamma_{ww}\,\delta_{z(a}\big)\partial_{b)}\tilde\gamma_{ww} \nonumber\\
&\qquad-\tfrac{1}{(2\tilde\gamma_{ww})^2}\,\partial_a\tilde\gamma_{ww}\,\partial_b\tilde\gamma_{ww}\Big\}\,,\\
&\tilde R_{ww}+2\,\tilde D_wZ_w = -\tfrac{1}{2}\,\tilde\gamma^{cd}\partial_c\partial_d\tilde\gamma_{ww}+\tfrac{1}{z}\,\tilde\gamma_{ww}\,\hat\Gamma^z \nonumber\\
&\quad-\tfrac{1}{2\tilde\gamma_{ww}}\Big(\tfrac{n}{z}\,\partial_z\tilde\gamma_{ww}+\tfrac{2}{z^2}\big(\tilde\gamma_{zz}-\tilde\gamma_{ww}\big)\Big) \nonumber\\
&\quad+\tfrac{1}{z^2\,\tilde\gamma_{ww}}\big(\tilde\gamma_{zz}-2\,\tilde\gamma_{ww}+\tilde\gamma^{zz}\,\tilde\gamma_{ww}^2\big)+\tfrac{1}{2}\,\hat\Gamma^a\partial_a\tilde\gamma_{ww} \nonumber\\
&\quad+\tfrac{1}{2\tilde\gamma_{ww}}\,\tilde\gamma^{ab}\,\partial_a\tilde\gamma_{ww}\,\partial_b\tilde\gamma_{ww}\,,\\
&R^\chi_{ab}+2\,D^\chi_{(a}Z_{b)} = \tfrac{d-2}{2\chi}\Big[\tilde D_a\tilde D_b\chi-\tfrac{1}{2\chi}\,\partial_a\chi\,\partial_b\chi\Big] \nonumber\\
&\quad+\tfrac{1}{2\chi}\,\tilde\gamma_{ab}\Big(\tilde D_i\tilde D^i\chi-\tfrac{d}{2\chi}\,\tilde\gamma^{cd}\,\partial_c\chi\,\partial_d\chi\Big) \nonumber\\
&\quad+\tfrac{2}{\chi}\Big(Z_{(a}\,\partial_{b)}\chi-\tfrac{1}{2}\,\tilde\gamma_{ab}\,\tilde\gamma^{cd}Z_c\,\partial_d\chi\Big)\,,\\
&R^\chi_{ww}+2\,D^\chi_wZ_w = -\tfrac{1}{\chi}\,\tilde\gamma_{ww}\,\tilde\gamma^{cd}Z_c\,\partial_d\chi \nonumber\\
&\quad+\tfrac{d-2}{2\chi}\Big[\tfrac{1}{z}\,\tilde\gamma_{ww}\,\tilde\gamma^{za}\partial_a\chi+\tfrac{1}{2}\,\tilde\gamma^{ab}\,\partial_a\chi\,\partial_b\tilde\gamma_{ww}\Big] \nonumber\\
&\quad+\tfrac{1}{2\chi}\,\tilde\gamma_{ww}\Big(\tilde D_i\tilde D^i\chi-\tfrac{d}{2\chi}\,\tilde\gamma^{cd}\,\partial_c\chi\,\partial_d\chi\Big)\,,
\end{align}
\end{subequations}
where $\tilde D_i\tilde D^i\chi$ is computed as in \eqref{eq:app_boxalpha} with $\alpha\to\chi$.

\paragraph{Riemann tensor and Gauss-Bonnet tensors.}
The Gauss-Bonnet terms require the full spatial Riemann tensor. Its non-vanishing cartoon components, in terms of the physical metric $\gamma_{ij}$ and its Christoffel symbols $\Gamma^i_{\phantom{i}jk}$ (with $\Gamma^a_{\phantom{a}ww}$ and $\Gamma^w_{\phantom{w}wa}$ given by \eqref{eq:app_Gamma_ww}--\eqref{eq:app_Gamma_wa} with $\tilde\gamma_{ij}\to\gamma_{ij}$), are
\begin{subequations}
\begin{align}
R_{abcd} =&~ \tfrac{1}{2}\big(\partial_d\partial_a\gamma_{bc}+\partial_c\partial_b\gamma_{ad} \nonumber\\
&-\partial_c\partial_a\gamma_{bd}-\partial_d\partial_b\gamma_{ac}\big) \nonumber\\
&-\gamma_{mn}\big(\Gamma^m_{\phantom{m}ac}\Gamma^n_{\phantom{n}bd}-\Gamma^m_{\phantom{m}ad}\Gamma^n_{\phantom{n}bc}\big)\,,\\
R_{aw_pbw_q} =&~ \delta_{pq}\,R_{awbw}\,,\nonumber\\
R_{awbw} =&~ -\tfrac{1}{2}\,\partial_a\partial_b\gamma_{ww}-\tfrac{1}{2z}\,\partial_z\gamma_{ab} \nonumber\\
&-\gamma_{cd}\,\Gamma^c_{\phantom{c}ab}\Gamma^d_{\phantom{d}ww}+\tfrac{1}{z}\big(\partial_{(a}\gamma_{b)z}-\delta_{z(a}\partial_{b)}\gamma_{ww}\big) \nonumber\\
&+\tfrac{1}{4}\,\gamma^{ww}\,\partial_a\gamma_{ww}\,\partial_b\gamma_{ww}\,,\\
R_{w_pw_qw_rw_s} =&~ \big(\delta_{pr}\delta_{qs}-\delta_{ps}\delta_{qr}\big)R_{wvwv}\,,\nonumber\\
R_{wvwv} =&~ -\tfrac{1}{4}\,\gamma^{ab}\,\partial_a\gamma_{ww}\,\partial_b\gamma_{ww}-\tfrac{1}{z}\,\gamma_{ww}\,\gamma^{za}\partial_a\gamma_{ww} \nonumber\\
&+\tfrac{1}{z^2}\,\gamma_{ww}\big(1-\gamma^{zz}\gamma_{ww}\big)\,,
\end{align}
\end{subequations}
where $v\neq w$ denotes a second sphere direction. From these,
\begin{subequations}
\begin{align}
R_{ab}&=\gamma^{cd}R_{acbd}+n\,\gamma^{ww}R_{awbw}\,,\\
R_{ww}&=\gamma^{ab}R_{awbw}+(n-1)\,\gamma^{ww}R_{wvwv}\,,
\end{align}
\end{subequations}
and $R=\gamma^{ab}R_{ab}+n\,\gamma^{ww}R_{ww}$.
The cartoon components of $M_{ijkl}$ and $M_{ij}$ are
\begin{subequations}
\begin{align}
M_{abcd} &= R_{abcd}+K_{ac}K_{bd}-K_{ad}K_{bc}\,, \\
M_{awbw} &= R_{awbw}+K_{ab}K_{ww}\,, \\
M_{wvwv} &= R_{wvwv}+K_{ww}^2\,, \\
M_{ab} &= R_{ab}+K\,K_{ab}-K_{ac}K^c_{~b}\,, \\
M_{ww} &= R_{ww}+K\,K_{ww}-\gamma^{ww}K_{ww}^2\,, \\
M &= R+K^2-K^{ab}K_{ab}-n\,(\gamma^{ww})^2K_{ww}^2\,,
\end{align}
\end{subequations}
with $M_{aw_pbw_q}=\delta_{pq}M_{awbw}$, $M_{w_pw_qw_rw_s}=(\delta_{pr}\delta_{qs}-\delta_{ps}\delta_{qr})M_{wvwv}$ and $M_{w_pw_q}=\delta_{pq}M_{ww}$. The non-vanishing components of $D_iK_{jk}$ are
\begin{subequations}
\begin{align}
D_aK_{bc} =&~ \partial_aK_{bc}-\Gamma^d_{\phantom{d}ab}K_{dc}-\Gamma^d_{\phantom{d}ac}K_{bd}\,,\\
D_aK_{w_pw_q} =&~ \delta_{pq}\,D_aK_{ww}\,, \nonumber\\
D_aK_{ww} =&~ \partial_aK_{ww}-2\,\Gamma^w_{\phantom{w}wa}K_{ww}\,,\\
D_{w_p}K_{w_qa} =&~ \delta_{pq}\,D_wK_{wa}\,, \nonumber\\
D_wK_{wa} =&~ \tfrac{1}{z}\big(K_{az}-\delta_{az}K_{ww}\big) \nonumber\\
&-\Gamma^b_{\phantom{b}ww}K_{ba}-\Gamma^w_{\phantom{w}wa}K_{ww}\,,
\end{align}
\end{subequations}
so that the non-vanishing components of $N_{ijk}$ and $N_i$ are
\begin{subequations}
\begin{align}
N_{abc}&=D_aK_{bc}-D_bK_{ac}\,,\\
N_{aw_pw_q}&=-N_{w_paw_q}=\delta_{pq}\,N_{aww}\,,\\
N_{aww}&=D_aK_{ww}-D_wK_{wa}\,,\\
N_a&=\gamma^{bc}N_{bac}-n\,\gamma^{ww}N_{aww}\,.
\end{align}
\end{subequations}

\paragraph{Coefficients of the $(\tilde A_{ij},K)$ system.}
The entries of the matrix in \eqref{eq:app_cartoon_AK} are
\begin{subequations}
\begin{align}
&X_{ab}^{\phantom{ab}cd} = \delta_a^{(c}\delta_b^{d)}+2\,\lambda^{\rm GB}\Big[M\,\delta_a^{(c}\delta_b^{d)}+\tfrac{6}{d}\,\gamma_{ab}M^{cd} \nonumber\\
&\qquad-2\big(M_{a\phantom{(c}b}^{\phantom{a}(c\phantom{b}d)}+2\,M_a^{\phantom{a}(c}\delta_b^{\phantom{b}d)}\big)\Big]\,,\\
&X_{ab}^{\phantom{ab}ww} = \lambda^{\rm GB}\,(\gamma^{ww})^2\Big[\tfrac{6}{d}\,\gamma_{ab}M_{ww}-2\,M_{awbw}\Big]\,,\\
&X_{ww}^{\phantom{ww}cd} = \lambda^{\rm GB}\Big[\tfrac{6}{d}\,\gamma_{ww}M^{cd}-2\,M_{w\phantom{c}w}^{\phantom{w}c\phantom{w}d}\Big]\,,\\
&X_{ww}^{\phantom{ww}ww} = 1+\lambda^{\rm GB}\Big[2\,M+\big(\tfrac{6}{d}-4\big)M_{ww}\,\gamma^{ww}\Big]\,,\\
&X^{cd} = -\tfrac{4(d-3)}{(d-1)\chi}\,\lambda^{\rm GB}M^{cd}\,,\\
&X^{ww} = -\tfrac{4(d-3)}{(d-1)\chi}\,\lambda^{\rm GB}M^{ww}\,,\\
&Y_{ab} = \tfrac{4(d-3)}{d}\,\lambda^{\rm GB}\chi\big(M_{ab}-\tfrac{1}{d}\,\gamma_{ab}M\big)\,,\\
&Y_{ww} = \tfrac{4(d-3)}{d}\,\lambda^{\rm GB}\chi\big(M_{ww}-\tfrac{1}{d}\,\gamma_{ww}M\big)\,,
\end{align}
\end{subequations}
and $Y=1+\tfrac{2(d-2)(d-3)}{d(d-1)}\,\lambda^{\rm GB}M$. The right-hand sides are
\begin{subequations}
\begin{align}
&Z_{ab} = {\mathcal L}_\beta\tilde A_{ab}+\chi\big[-D_aD_b\alpha+\alpha\big(R_{ab}+2\,D_{(a}Z_{b)}\big)\big]^{\text{TF}} \nonumber\\
&\qquad+\alpha\big[\tilde A_{ab}(K-2\Theta)-2\,\tilde A_{ac}\tilde A^c_{~b}\big] \nonumber\\
&\qquad-\tfrac{2}{d}\,\text{div}\beta\,\tilde A_{ab}+\lambda^{\rm GB}\,\tilde S^{GB}_{ab}\,,\\
&Z_{ww} = {\mathcal L}_\beta\tilde A_{ww}+\chi\big[-D_wD_w\alpha+\alpha\big(R_{ww}+2\,D_wZ_w\big)\big]^{\text{TF}} \nonumber\\
&\qquad+\alpha\big[\tilde A_{ww}(K-2\Theta)-2\,\tilde\gamma^{ww}\tilde A_{ww}^2\big] \nonumber\\
&\qquad-\tfrac{2}{d}\,\text{div}\beta\,\tilde A_{ww}+\lambda^{\rm GB}\,\tilde S^{GB}_{ww}\,,\\
&Z={\mathcal L}_{\beta}K-D^iD_i\alpha+\alpha\big[R+2\,D_iZ^i+K(K-2\Theta)\big] \nonumber\\
&\qquad-d\,\kappa_1(1+\kappa_2)\,\alpha\,\Theta+\lambda^{\rm GB}\,S^{GB}_K \nonumber\\
&\qquad-\tfrac{d\,b(x)}{2(d-1)(1+b(x))}\,\alpha\Big[H-\tfrac{2(d-1+(d-2)b(x))}{d-1}\,\tfrac{1}{\alpha}\,Z_cD^c\alpha \nonumber\\
&\qquad\quad-(d-1)\kappa_1(2+\kappa_2)\,\Theta-2\,\lambda^{\rm GB}\rho^{GB}\Big]\,,
\end{align}
\end{subequations}
with ${\mathcal L}_\beta\tilde A_{ab}=\beta^c\partial_c\tilde A_{ab}+2\,\tilde A_{c(a}\partial_{b)}\beta^c$, ${\mathcal L}_\beta\tilde A_{ww}=\beta^c\partial_c\tilde A_{ww}+\tfrac{2}{z}\,\beta^z\tilde A_{ww}$ and ${\mathcal L}_\beta K=\beta^c\partial_cK$.

\paragraph{Source terms.}
The Gauss-Bonnet source terms are built from the cartoon components of $\hat{\mathcal O}_{ij}$ and $\hat{\mathcal O}$,
\begin{subequations}
\begin{align}
\hat{\mathcal O}_{ab} =&~ {\mathcal L}_\beta\tilde A_{ab}-\alpha\big(\tilde A_{ac}\tilde A_b^{~c}-\tfrac{2}{d}\,K\tilde A_{ab}-\tfrac{1}{d^2}\,K^2\tilde\gamma_{ab}\big) \nonumber\\
&-\chi\,D_aD_b\alpha-\tfrac{2}{d}\,\tilde A_{ab}\,\text{div}\beta+\tfrac{1}{d}\,\tilde\gamma_{ab}\,{\mathcal L}_\beta K\,,\\
\hat{\mathcal O}_{ww} =&~ {\mathcal L}_\beta\tilde A_{ww}-\alpha\big(\tilde\gamma^{ww}\tilde A_{ww}^2-\tfrac{2}{d}\,K\tilde A_{ww} \nonumber\\
&-\tfrac{1}{d^2}\,K^2\tilde\gamma_{ww}\big)-\chi\,D_wD_w\alpha \nonumber\\
&-\tfrac{2}{d}\,\tilde A_{ww}\,\text{div}\beta+\tfrac{1}{d}\,\tilde\gamma_{ww}\,{\mathcal L}_\beta K\,,\\
\hat{\mathcal O} =&~ {\mathcal L}_\beta K-D_iD^i\alpha \nonumber\\
&+\alpha\big(\tilde A_{ab}\tilde A^{ab}+n\,\tilde A_{ww}\tilde A^{ww}+\tfrac{1}{d}\,K^2\big)\,,
\end{align}
\end{subequations}
and read
\begin{subequations}
\begin{align}
&\rho^{GB} = -\tfrac{1}{2}\Big[M^2-4\big(M_{ab}M^{ab}+n\,M_{ww}M^{ww}\big) \nonumber\\
&\qquad+M_{abcd}M^{abcd}+4\,n\,M_{awbw}M^{awbw} \nonumber\\
&\qquad+2\,n(n-1)\,M_{wvwv}M^{wvwv}\Big]\,,\\
&J^{GB}_a = -2\Big[M N_a-2\,M_a^{~b}N_b \nonumber\\
&\qquad+2\big(M^{bc}N_{abc}+n\,M^{ww}N_{aww}\big) \nonumber\\
&\qquad-\big(M_a^{~bcd}N_{cdb}+2\,n\,M_a^{~wbw}N_{bww}\big)\Big]\,,\\
&\tilde S^{GB}_{ab} = 4\,\chi\,M_{ab}\hat{\mathcal O}+2\,M\hat{\mathcal O}_{ab}-8\,M_{(a}^{\phantom{(a}c}\hat{\mathcal O}_{b)c} \nonumber\\
&\quad+\tfrac{12}{d}\,\tilde\gamma_{ab}\,\hat S^{GB}-4\big(M_{a\phantom{c}b}^{\phantom{a}c\phantom{b}d}\hat{\mathcal O}_{cd}+n\,M_{a\phantom{w}b}^{\phantom{a}w\phantom{b}w}\hat{\mathcal O}_{ww}\big) \nonumber\\
&\quad+2\,\alpha\,\chi\Big[M M_{ab}-2\big(M_{ac}M^c_{~b}+M^{cd}M_{acbd} \nonumber\\
&\qquad+n\,M^{ww}M_{awbw}\big)+M_{acde}M_b^{\phantom{b}cde} \nonumber\\
&\qquad+2\,n\,M_{awcw}M_b^{\phantom{b}wcw}+2\big(N_aN_b-2\,N^cN_{c(ab)} \nonumber\\
&\qquad-\tfrac{1}{2}\,N_{cda}N^{cd}_{\phantom{cd}b}-N_{acd}N_b^{\phantom{b}cd} \nonumber\\
&\qquad-n\,N_{aww}N_b^{\phantom{b}ww}\big)\Big]\,,\\
&\tilde S^{GB}_{ww} = 4\,\chi\,M_{ww}\hat{\mathcal O}+2\,M\hat{\mathcal O}_{ww}-8\,M_w^{~w}\hat{\mathcal O}_{ww} \nonumber\\
&\quad+\tfrac{12}{d}\,\tilde\gamma_{ww}\,\hat S^{GB} \nonumber\\
&\quad-4\big[M_{w\phantom{a}w}^{\phantom{w}a\phantom{w}b}\hat{\mathcal O}_{ab}+(n-1)\,M_{wvwv}\,\hat{\mathcal O}^{ww}\big] \nonumber\\
&\quad+2\,\alpha\,\chi\Big[M M_{ww}-2\big(M_{ww}M^w_{~w}+M^{ab}M_{awbw} \nonumber\\
&\qquad+(n-1)\,M^{ww}M_{wvwv}\big)+2\,M_{wawb}M_w^{\phantom{w}awb} \nonumber\\
&\qquad+2(n-1)\,M_{wvwv}M_w^{\phantom{w}vwv} \nonumber\\
&\qquad-4\big(N^aN_{aww}+N_{aww}N^{aw}_{\phantom{aw}w}\big)\Big]\,,\\
&S^{GB}_K = \tfrac{1}{d-1}\Big\{2(d-3)\Big[M\hat{\mathcal O} \nonumber\\
&\qquad-\tfrac{2}{\chi}\big(M^{ab}\hat{\mathcal O}_{ab}+n\,M^{ww}\hat{\mathcal O}_{ww}\big) \nonumber\\
&\qquad+\alpha\big(2\,N_aN^a-N_{abc}N^{abc} \nonumber\\
&\qquad-2\,n\,N_{aww}N^{aww}\big)\Big]-\alpha(d-4)\,\rho^{GB}\Big\}\,,
\end{align}
\end{subequations}
where
\begin{align}
\hat S^{GB}\equiv&-\tfrac{1}{2}\,M\hat{\mathcal O}+\tfrac{1}{\chi}\big(M^{cd}\hat{\mathcal O}_{cd}+n\,M^{ww}\hat{\mathcal O}_{ww}\big) \nonumber\\
&+\tfrac{\alpha}{3}\,\rho^{GB}-\alpha\Big[N_aN^a \nonumber\\
&-\tfrac{1}{2}\big(N_{abc}N^{abc}+2\,n\,N_{aww}N^{aww}\big)\Big]\,,
\end{align}
and indices along the sphere are raised with $\gamma^{ww}=\gamma^{vv}$, e.g.\ $M_w^{\phantom{w}vwv}=M_{wvwv}(\gamma^{ww})^3$ and $\hat{\mathcal O}^{ww}=(\gamma^{ww})^2\,\hat{\mathcal O}_{ww}$.

\clearpage 
\bibliography{reference}
\end{document}